\documentclass[aps,prb,reprint,longbibliography,superscriptaddress,citeautoscript,nofootinbib,floatfix]{revtex4-2}
\usepackage[utf8]{inputenc}
\usepackage[T1]{fontenc}

\usepackage[american]{babel}

\usepackage{microtype}
\usepackage{amsmath}
\usepackage[normalem]{ulem}

\usepackage{graphicx}
\usepackage{float}
\usepackage{xcolor}
\usepackage{siunitx}
\usepackage{tikz}

\newcommand{\mytitle}{Universality of grain boundary phases in fcc
  metals:\\ Case study on high-angle [111] symmetric tilt grain
  boundaries}

\usepackage{hyperref} %
\hypersetup{
  final,
  pdfborder={0 0 0},
  pdftitle={\mytitle},
  pdfauthor={Brink, Langenohl, Bishara, Dehm},
  colorlinks=true,
  urlcolor=blue,
  linkcolor=blue,
  citecolor=blue
}

\begin{document}
\frenchspacing

\title{\mytitle}

\author{Tobias Brink}
\email{t.brink@mpie.de}
\affiliation{Max-Planck-Institut f\"ur Eisenforschung GmbH,
  Max-Planck-Stra\ss{}e 1, 40237 D\"usseldorf, Germany}

\author{Lena Langenohl}
\affiliation{Max-Planck-Institut f\"ur Eisenforschung GmbH,
  Max-Planck-Stra\ss{}e 1, 40237 D\"usseldorf, Germany}

\author{Hanna Bishara}
\affiliation{Max-Planck-Institut f\"ur Eisenforschung GmbH,
  Max-Planck-Stra\ss{}e 1, 40237 D\"usseldorf, Germany}

\author{Gerhard Dehm}
\affiliation{Max-Planck-Institut f\"ur Eisenforschung GmbH,
  Max-Planck-Stra\ss{}e 1, 40237 D\"usseldorf, Germany}

\date{\today}

\begin{abstract}
  Grain boundaries often exhibit ordered atomic structures. Increasing
  amounts of evidence have been provided by transmission electron
  microscopy and atomistic computer simulations that different stable
  and metastable grain boundary structures can occur. Meanwhile,
  theories to treat them thermodynamically as grain boundary phases
  have been developed. Whereas atomic structures were identified at
  particular grain boundaries for particular materials, it remains an
  open question if these structures and their thermodynamic excess
  properties are material specific or generalizable to, e.g., all fcc
  metals. In order to elucidate that question, we use atomistic
  simulations with classical interatomic potentials to investigate a
  range of high-angle $[11\overline{1}]$ symmetric tilt grain
  boundaries in Ni, Cu, Pd, Ag, Au, Al, and Pb. We could indeed find
  two families of grain boundary phases in all of the investigated
  grain boundaries, which cover most of the standard fcc materials.
  Where possible, we compared the atomic structures to
  atomic-resolution electron microscopy images and found that the
  structures match. This poses the question if the grain boundary
  phases are simply the result of sphere-packing geometry or if
  material-specific bonding physics play a role. We tested this using
  simple model pair potentials and found that medium-ranged
  interactions are required to reproduce the atomic structures, while
  the more realistic material models mostly affect the grain boundary
  (free) energy. In addition to the structural investigation, we also
  report the thermodynamic excess properties of the grain boundaries,
  explore how they influence the thermodynamic stability of the grain
  boundary phases, and detail the commonalities and differences
  between the materials.
\end{abstract}

\maketitle

\newcounter{supplfigctr}
\renewcommand{\thesupplfigctr}{\arabic{supplfigctr}}
{\refstepcounter{supplfigctr}\label{fig:finding-B}}
{\refstepcounter{supplfigctr}\label{fig:snapshots-S13b}}
{\refstepcounter{supplfigctr}\label{fig:snapshots-S7}}
{\refstepcounter{supplfigctr}\label{fig:snapshots-S49}}
{\refstepcounter{supplfigctr}\label{fig:snapshots-S19b}}
{\refstepcounter{supplfigctr}\label{fig:snapshots-S37c}}
{\refstepcounter{supplfigctr}\label{fig:STEM-S49-full}}
{\refstepcounter{supplfigctr}\label{fig:excess-S13b}}
{\refstepcounter{supplfigctr}\label{fig:excess-S7}}
{\refstepcounter{supplfigctr}\label{fig:excess-S49}}
{\refstepcounter{supplfigctr}\label{fig:excess-S19b}}
{\refstepcounter{supplfigctr}\label{fig:excess-S37c}}
{\refstepcounter{supplfigctr}\label{fig:gb-energy}}
{\refstepcounter{supplfigctr}\label{fig:correlation-SFE-nonorm}}
{\refstepcounter{supplfigctr}\label{fig:sigma33-S13b}}
{\refstepcounter{supplfigctr}\label{fig:sigma33-S7}}
{\refstepcounter{supplfigctr}\label{fig:sigma33-S49}}
{\refstepcounter{supplfigctr}\label{fig:sigma33-S19b}}
{\refstepcounter{supplfigctr}\label{fig:sigma33-S37c}}
{\refstepcounter{supplfigctr}\label{fig:md-sigma33-S49-Pd}}
{\refstepcounter{supplfigctr}\label{fig:epsilon22-S13b}}
{\refstepcounter{supplfigctr}\label{fig:epsilon22-S7}}
{\refstepcounter{supplfigctr}\label{fig:epsilon22-S49}}
{\refstepcounter{supplfigctr}\label{fig:epsilon22-S19b}}
{\refstepcounter{supplfigctr}\label{fig:epsilon22-S37c}}
{\refstepcounter{supplfigctr}\label{fig:md-epsilon22-S7-Cu}}
{\refstepcounter{supplfigctr}\label{fig:temperature-S13b}}
{\refstepcounter{supplfigctr}\label{fig:temperature-S7}}
{\refstepcounter{supplfigctr}\label{fig:temperature-S49}}
{\refstepcounter{supplfigctr}\label{fig:temperature-S19b}}
{\refstepcounter{supplfigctr}\label{fig:temperature-S37c}}
{\refstepcounter{supplfigctr}\label{fig:md-temperature-S7-Ni}}
{\refstepcounter{supplfigctr}\label{fig:md-temperature-S7-Cu}}
{\refstepcounter{supplfigctr}\label{fig:md-temperature-S7-Pd}}
{\refstepcounter{supplfigctr}\label{fig:md-temperature-S7-Ag}}
{\refstepcounter{supplfigctr}\label{fig:md-temperature-S7-Al}}
{\refstepcounter{supplfigctr}\label{fig:md-temperature-S37c-Ni}}
{\refstepcounter{supplfigctr}\label{fig:Al-pot}}
{\refstepcounter{supplfigctr}\label{fig:Pd-Au-pots}}
\section{Introduction}

\begin{tikzpicture}[remember picture,overlay]
  \node [anchor=north west, font=\footnotesize, align=left,
         text width=7.05in, xshift=0.75in, yshift=0.75in, inner xsep=0pt]
        at (current page.south west)
        {Published in:\\
         \href{https://doi.org/10.1103/PhysRevB.107.054103}
              {T.~Brink et al., Phys.~Rev.~B~\textbf{107}, 054103 (2023)}
         \hfill
         DOI: \href{https://doi.org/10.1103/PhysRevB.107.054103}
                   {10.1103/PhysRevB.107.054103}};
\end{tikzpicture}%
Grain boundaries (GBs) are defined by five macroscopic degrees of
freedom, describing the misorientation of the abutting crystallites
and the GB plane \cite{Priester2013}. At the atomic scale, GBs have
additional microscopic degrees of freedom \cite{Priester2013}, meaning
that a GB with a specific misorientation and GB plane can exhibit
different atomic structures. These distinct structures are called GB
phases \cite{Frolov2015a} or complexions \cite{Tang2006, Dillon2007,
  Cantwell2014, Cantwell2020} in analogy to bulk phases, because they
can be understood using a thermodynamic framework \cite{GibbsVol1,
  Frolov2012,Frolov2012a, Kaplan2013, Cantwell2014, Cantwell2020}. It
should be noted, however, that GB phases are not the same as bulk
phases insofar they can only exist at interfaces and not on their own,
in contrast to, e.g., bulk wetting phases or precipitates, which can
also appear at GBs \cite{Kaplan2013, Cantwell2020}.  GB phase
transitions have been proposed theoretically already from the 1960s
onwards \cite{Hart1968, Cahn1982, Rottman1988}. While such transitions
can be driven by segregation, as for example in
Refs.~\cite{Ference1988, Sigle2002, Dillon2007, Frolov2015,
  Khalajhedayati2015, Pan2016, Peter2018}, they also occur in pure
materials as demonstrated with atomistic simulations \cite{Wu2009,
  Frolov2013, Hickman2017, Aramfard2018, Zhu2018a, Frolov2018a,
  Frolov2018, Yang2020, Meiners2020, Frommeyer2022}, and
experimentally by atomic-resolution (scanning) transmission electron
microscopy (TEM, STEM) \cite{Mills1992, Mills1993, Meiners2020,
  Frommeyer2022}. Experimental observation, however, remains difficult
because at least two GB phases have to exist in stable or metastable
states under experimental conditions, which appears to be
rare. Available data suggest that GB phase transitions may influence
diffusivity \cite{Rabkin1999, Divinski2012, Frolov2013a,
  Rajeshwari2020}, GB motion \cite{Frolov2014, Wei2021}, intergranular
fracture \cite{Luo2011, Pan2015, Feng2018}, and electrical
conductivity \cite{Bishara2021}, among other material properties.

In the case of pure metals, computer simulations have been performed
predominantly to demonstrate the existence of different GB phases for
example cases, such as for specific macroscopic degrees of freedom or
for a single material. Comprehensive studies have been attempted in
the 1970s and 1980s, e.g., for tilt GBs of Al and Cu
\cite{Sutton1983}, but were limited by the use of simple pair
potentials and by mostly disregarding metastable GB structures. More
recently, GB phase transitions have been simulated in a $\Sigma5$ GB
for Cu, Ag, Au, and Ni \cite{Frolov2013}; in Cu for different
misorientations of $[001]$ tilt GBs \cite{Hickman2017, Aramfard2018,
  Zhu2018a, Yang2020}; in W for a variety of tilt GBs
\cite{Frolov2018a, Frolov2018, Yang2020}; and in Mg tilt GBs
\cite{Yang2020}. Apart from the early simulations on $\Sigma5$ GBs
\cite{Frolov2013}, where the same motifs were found for all metals, it
was not investigated if specific GB phases and their transitions are
generalizable to all fcc, bcc, or hcp materials, respectively. Thus it
is not clear to what extend GB phases are influenced by, e.g.,
bonding, structure, or packing density and if they can be correlated
with bulk material properties.

It has, however, long been known for tilt GBs that certain structural
motifs exist over a range of misorientations, leading to the
development of the structural unit model \cite{Bishop1968, Sutton1983,
  Sutton1983a, Sutton1983b}. This model describes GB structures as
combinations of motifs resulting from certain delimiting boundaries,
which are the GBs containing only a single motif and which serve
as reference structures for general GBs. A weakness of the model is
the assumption of a single, canonical structure for the delimiting
boundaries, which is in opposition to the existence of metastable GB
phases. Indeed, different ground states were for example found in
copper for the closely-related $\Sigma19$b and $\Sigma37$c tilt GBs
\cite{Meiners2020, Frommeyer2022}, which make the original strucural
unit model inapplicable. These shortcomings were addressed by the
development of a revised model that includes multiple---possibly
metastable---motifs for the delimiting boundaries \cite{Han2017}. In
atomistic simulations on tungsten \cite{Han2017}, this model
demonstrates on one hand that the varying motifs and their
combinations lead to different stable GB phases with varying
misorientations. On the other hand, it reaffirms that the motifs
remain existent across variations of the macroscopic degrees of
freedom of the GBs, even if only in a metastable state.

Nevertheless, systematic and quantitative studies among comparable
materials are uncommon and the question remains if specific GB phases
are universal features (for example of a given lattice structure of
the bulk crystal) or very specific to a material. To that end, the
present work is dedicated to atomistic computer simulations of
symmetric $[11\overline{1}]$ tilt GBs in a range of fcc metals and
expands on recent results for $\Sigma19$b and $\Sigma37$c tilt GBs in
copper \cite{Meiners2020, Frommeyer2022}. While some experimental data
regarding the atomic structure of the GB phases is available for Cu
and Al \cite{Meiners2020, Frommeyer2022, SabaZoo}, we will expand the
computer investigation to most of the fcc metals for which reasonable
interatomic potentials are available and to a range of
misorientations. In addition, we use pair potentials to switch off
environment-dependent bond energies (bond order) and/or medium-ranged
interatomic interactions beyond the first neighbor shell. This
tunability allows us to study if the atomic structures of the GB
phases are defined more by packing geometry or the material-specific
physics of bonding. We present some common trends and differences
between the materials.

\section{Methods}

We modeled GBs in bicrystals using embedded atom method (EAM)
potentials for Ni \cite{Mishin2004}, Cu \cite{Mishin2001}, Pd
\cite{Foiles2001}, Ag \cite{Williams2006}, Au \cite{Foiles1986}, and
Al \cite{Mishin1999}, as well as a modified EAM (MEAM) potential for
Pb \cite{Lee2003}. The potential files were downloaded from the NIST
Interatomic Potentials Repository \cite{NIST-IPR}, except for the Pd
potential, which we reproduced from the data in the original
publication, and the Al potential, which was smoothed as described in
Appendix~\ref{sec:numerical-issues} to overcome numerical problems
with free energy calculations. Molecular statics and molecular
dynamics (MD) simulations were performed using the software
\textsc{lammps} \cite{Plimpton1995, Thompson2022}.

In addition to the more realistic potentials, we also used generic
model pair potentials to evaluate how much material-specific physics
is required to reproduce the results of the (M)EAM potentials. For
these, we use reduced units in terms of the equilibrium bond length
$r_0$ in fcc and the corresponding fcc cohesive energy $E_\text{coh}$
per atom (which is by the convention used for interatomic potentials
related to the total energy in the ground state via
$E_\text{coh} = -E_0^\text{fcc}$). Here, we considered a Lennard-Jones
potential with a cutoff of $2.5r_0$ (corresponding to 6 fcc
next-neighbor shells), shifted so that the bond energy at the cutoff
is zero. The parameters $\sigma_\text{LJ} = 0.91303 r_0$ and
$\varepsilon_\text{LJ} = 0.12927915 E_\text{coh}$ were chosen to
obtain a lattice constant of $a_0^\text{fcc} = \sqrt{2} r_0$ and
$E_0^\text{fcc} = -1 E_\text{coh}$. This parametrization has a stable
fcc and metastable hcp phase (energy difference of roughly
$0.001E_\text{coh}$).

Furthermore, in order to investigate the difference between
medium-ranged and next-neighbor-only interactions, we constructed pair
potentials of the form
\begin{equation}
  \label{eq:nn-pp}
  E_i = \sum_{j\neq i}
  \begin{cases}
    C \left[
      \left(
        \frac{\sigma_\text{pp}}{r_{ij}}
      \right)^{n_\text{pp}}
      \hspace{-0.5ex} -
      \left(
        \frac{\sigma_\text{pp}}{r_{ij}}
      \right)^{m_\text{pp}}
    \right]
    & r \leq R_\text{inner} \\[8pt]
    \sum_{k=0}^4 c_k^\text{pp} r^k
    & R_\text{inner} < r \leq R_\text{cut} \\[8pt]
    0 & r > R_\text{cut},
  \end{cases}
\end{equation}
with $R_\text{cut} = 1.35r_0$, so that only the first fcc neighbor
shell is included. It is
\begin{equation}
  C = \varepsilon_\text{pp} n_\text{pp} (n_\text{pp}-m_\text{pp})^{-1}
  \left(\frac{n_\text{pp}}{m_\text{pp}}\right)^\frac{m_\text{pp}}{(n_\text{pp}-m_\text{pp})},
\end{equation}
and we used $n_\text{pp} = 24$, $m_\text{pp} = 14,16,18,20,22$.  This
leads to bond stiffnesses that are higher than the standard
Lennard-Jones potential ($n_\text{pp} = 12$, $m_\text{pp} = 6$), but
this is required to obtain a resonably-shaped potential well inside
the very short cutoff range. The polynomial with $c_0^\text{pp}$,
$c_1^\text{pp}$, $c_2^\text{pp}$, $c_3^\text{pp}$, $c_4^\text{pp}$ is
defined so that the potentials are continuous up to the second
derivative at $R_\text{inner}$ and both energy and force are zero at
the cutoff $R_\text{cut}$. We used $R_\text{inner} = 1.1r_0$. We
defined five different potentials (different $m_\text{pp}$) to see if
the bond stiffness influences the GB structures, but found that this
is not the case here. We consequently report only the results of the
potential with $m_\text{pp} = 18$ in the rest of this paper. All
potential files are available in the companion dataset
\cite{Brink2022zenodo}.

\subsection{Bulk properties of fcc crystals and evaluation of the
  potentials}

\begin{table*}
    \centering
    \caption{Material properties of fcc transition metals computed
      with the EAM potentials (pot.) compared to literature values
      (ref.).
      We list experimental ground-state energies energies
      $E_0^\text{fcc}$ from Ref.~\cite{Kittel2005}; experimental
      lattice constants $a_0^\text{fcc}$, elastic constants $c_{ij}$,
      bulk modulus $K$, and melting points $T_\text{melt}$ from
      Ref.~\cite{CRCHandbook2022}; experimental vacancy formation
      energies $E_{f,\text{vac}}$ from Ref.~\cite{Ullmaier1991} (note
      that the Pd data is only from a single measurement and less
      reliable than the other data points); surface energies
      $\gamma_{(111)}$ from DFT calculations for the (111) surface
      that match the average experimental values reasonably well
      \cite{Vitos1998}; experimental stacking-fault energies
      $\gamma_\text{SF}$ and DFT values for the unstable
      stacking-fault energies $\gamma_\text{USF}$ as collected in the
      literature review in Ref.~\cite{Bernstein2004}; and DFT
      calculations of the maximum shear stress $\tau_\text{SF}$ along
      the generalized stacking-fault curve from
      Ref.~\cite{Ogata2002}.}
    \label{tab:mater-prop-transition}
    \begin{ruledtabular}
    \begin{tabular}{rcccccccccc}
                                       & \multicolumn{2}{c}{Ni} & \multicolumn{2}{c}{Cu}            & \multicolumn{2}{c}{Pd}        & \multicolumn{2}{c}{Ag}            & \multicolumn{2}{c}{Au}           \\
                                       & pot.     & ref.        & pot.              & ref.          & pot.          & ref.          & pot.              & ref.          & pot.             & ref.          \\
        \colrule
        $E_0^\text{fcc}$ (eV/atom)     & $-4.450$ & $-4.44$     & $-3.540$          & $-3.49$       & $-3.910$      & $-3.89$       & $-2.850$          & $-2.95$       & $-3.930$         & $-3.81$       \\
        $a_0^\text{fcc}$ (\AA)         & $3.520$  & $3.524$     & $3.615$           & $3.615$       & $3.890$       & $3.890$       & $4.090$           & $4.086$       & $4.080$          & $4.078$       \\
        $E_0^\text{hcp}$ (eV/atom)     & $-4.428$ &             & $-3.532$          &               & $-3.878$      &               & $-2.846$          &               & $-3.929$         &               \\
        $a_0^\text{hcp}$ (\AA)         & $2.482$  &             & $2.556$           &               & $2.719$       &               & $2.896$           &               & $2.886$          &               \\
        $c_0^\text{hcp}$ (\AA)         & $4.105$  &             & $4.162$           &               & $4.675$       &               & $4.679$           &               & $4.704$          &               \\
        $c_{11}$ (GPa)                 & 241      & 248         & 170               & 168           & 239           & 227           & 124               & 124           & 183              & 192           \\
        $c_{12}$ (GPa)                 & 151      & 155         & 123               & 122           & 173           & 176           & \phantom{0}94     & \phantom{0}94 & 159              & 163           \\
        $c_{44}$ (GPa)                 & 127      & 124         & \phantom{0}76     & \phantom{0}76 & \phantom{0}66 & \phantom{0}72 & \phantom{0}46     & \phantom{0}46 & \phantom{0}45    & \phantom{0}42 \\
        $K$ (GPa)                      & 181      & 186         & 138               & 138           & 195           & 193           & 104               & 104           & 167              & 173           \\
        $T_\text{melt}$ (K)            & 1698     & 1728        & 1324              & 1358          & 1154          & 1828          & 1266              & 1235          & 1111             & 1337          \\
        $E_{f,\text{vac}}$ (eV)        & $1.571$  & 1.79        & $1.272$           & 1.28          & $1.375$       & 1.7           & $1.103$           & 1.11          & $1.026$          & 0.93          \\
        $\gamma_{(111)}$ (J/m$^2$)     & $1.759$  & $2.011$     & $1.239$           & $1.952$       & $1.922$       & $1.920$       & $0.862$           & $1.172$       & $0.786$          & $1.283$       \\
        $\gamma_\text{SF}$ (mJ/m$^2$)  & $134.7$  & 125--300    & $\phantom{0}44.4$ & 35--78        & $181.1$       & 175--180      & $\phantom{0}17.8$ & 16--22        & $\phantom{0}4.7$ & 30--45        \\
        $\gamma_\text{USF}$ (mJ/m$^2$) & $297.6$  & 269--350    & $162.1$           & 158--210      & $211.7$       & 265           & $114.9$           & 190           & $95.7$           &               \\
        $\tau_\text{SF}$ (GPa)         & $5.8$    &             & $3.2$             & $2.2$         & $3.4$         &               & $2.0$             &               & $1.8$            &               \\
    \end{tabular}
    \end{ruledtabular}
\end{table*}

\begin{table}
    \centering
    \caption{Material properties of other fcc metals computed with the
      (M)EAM potentials (pot.) compared to literature values
      (ref.). Reference data sources are the same as in
      Table~\ref{tab:mater-prop-transition}.}
    \label{tab:mater-prop-other}
    \begin{ruledtabular}
    \begin{tabular}{rcccc}
                                       & \multicolumn{2}{c}{Al}        & \multicolumn{2}{c}{Pb}         \\
                                       & pot.          & ref.          & pot.             & ref.        \\
        \colrule
        $E_0^\text{fcc}$ (eV/atom)     & $-3.360$      & $-3.39$       & $-2.040$         & $-2.03$     \\
        $a_0^\text{fcc}$ (\AA)         & $4.050$       & $4.050$       & $4.950$          & $4.950$     \\
        $E_0^\text{hcp}$ (eV/atom)     & $-3.332$      &               & $-2.037$         &             \\
        $a_0^\text{hcp}$ (\AA)         & $2.819$       &               & $3.497$          &             \\
        $c_0^\text{hcp}$ (\AA)         & $4.945$       &               & $5.729$          &             \\
        $c_{11}$ (GPa)                 & 114           & 107           & 56               & 50          \\
        $c_{12}$ (GPa)                 & \phantom{0}62 & \phantom{0}60 & 45               & 42          \\
        $c_{44}$ (GPa)                 & \phantom{0}32 & \phantom{0}28 & 19               & 15          \\
        $K$ (GPa)                      & 79            & \phantom{0}76 & 49               & 45          \\
        $T_\text{melt}$ (K)            & 1042          & 933           & 686              & 601         \\
        $E_{f,\text{vac}}$ (eV)        & $0.675$       & 0.67          & $0.584$          & 0.58        \\
        $\gamma_{(111)}$ (J/m$^2$)     & $0.871$       & $1.199$       & $0.362$          & $0.321$     \\
        $\gamma_\text{SF}$ (mJ/m$^2$)  & $145.5$       & 135--200      & $\phantom{0}9.0$ & 25          \\
        $\gamma_\text{USF}$ (mJ/m$^2$) & $167.3$       & 175--224      & $57.1$           &             \\
        $\tau_\text{SF}$ (GPa)         & $2.3$         & 2.8           &                  &             \\
    \end{tabular}
    \end{ruledtabular}
\end{table}

In order to evaluate the performance of the (M)EAM potentials, we
first computed the properties of the bulk fcc and hcp phases (listed
in Tables~\ref{tab:mater-prop-transition} and
\ref{tab:mater-prop-other}). Ground state energies $E_0$ (which by
convention are related to the cohesive energy via
$E_\text{coh} = -E_0^\text{fcc}$) and lattice constants $a_0$, $c_0$
at temperature $T = \SI{0}{K}$ were calculated using molecular statics
calculations on defect-free fcc systems, while vacancy formation
energies $E_{f,\text{vac}}$, (111) surface energies $\gamma_{(111)}$,
and stacking-fault energies $\gamma_\text{SF}$ were calculated using
systems containing the relevant defects. Generalized stacking-fault
curves were computed using the procedure described in
Ref.~\cite{Ogata2002}. Unstable stacking-fault energies
$\gamma_\text{USF}$ as well as the maximum shear stresses
$\tau_\text{SF}$ along the stacking-fault curves are reported in
Tables~\ref{tab:mater-prop-transition} and
\ref{tab:mater-prop-other}. We computed the elastic constants $c_{ij}$
and the bulk modulus $K$ for the fcc phase using the scripts
distributed with \textsc{lammps}, which derive the stiffness tensor
from the stress tensor by systematically applying strains to a
periodic fcc cell.

Melting points were computed with the method and software from
Ref.~\cite{Zhu2021}, which uses the interface method, i.e., a
crystal/liquid interface is constructed and simulated at different
temperatures with MD. The movement of the interface is monitored to
estimate the melting point.

Finally, we simulated the thermal expansion of the metals by measuring
the lattice constant at different temperatures using MD simulations.
We used the careful procedure from Ref.~\cite{Freitas2016} to achieve
high accuracy: Using a small timestep of $\delta t = \SI{0.5}{fs}$, a
barostat at \SI{0}{Pa} with damping parameter \SI{0.5}{ps}, and a
Langevin thermostat with a damping parameter of \SI{0.05}{ps} set to
maintain a total force of zero, we equilibrated a simulation cell
consisting of $20\times20\times20$ unit cells in \SI{50}{K} increments
up to the melting point at each temperature for
\SI{250}{ps}. Averaging of the lattice constant was performed over the
last \SI{60}{ps}. The corresponding raw data of the bulk property
calculations is available in the companion dataset
\cite{Brink2022zenodo}.

The data provided in Tables~\ref{tab:mater-prop-transition} and
\ref{tab:mater-prop-other} suggest that the Ni \cite{Mishin2004}, Cu
\cite{Mishin2001}, Ag \cite{Williams2006}, Al \cite{Mishin1999}, and
Pb \cite{Lee2003} potentials reproduce the bulk properties well. The
Pd potential \cite{Foiles2001} strongly underestimates the melting
point, while the Au potential \cite{Foiles1986} both underestimates
the melting point and the stacking-fault energy. The latter should
therefore be treated as a model potential for the case of a very low
stacking-fault energy. Surface energies are not predicted well in
general, but should not affect the simulation of GBs. The Ni, Cu, and
Ag potentials are based on closed-form expressions that vary smoothly
and continuously as a function of, e.g., bond length. Apart from bulk
and defect properties, these potentials were also tested to reproduce
thermal expansion and phonon frequencies, which are important for the
GB excess free energy calculations. The Al potential was produced with
similar care, but it was defined in terms of cubic splines. This can
lead to different behavior in different ranges of bond lengths, which
manifests for example in the GB free energy as shown later. The Au and
Pd potentials were defined to exactly follow an equation of state,
which often leads to inferior results \cite{Mishin2001}. The more
well-known, older Pd potential by Foiles et al.\ \cite{Foiles1986} has
been found to perform worse, with the newer potential used here
\cite{Foiles2001} reproducing the bulk material properties and
stacking-fault energies quite well \cite{Stukowski2010c,
  Stukowski2010PhD}. There exists to the best of our knowledge no
reasonable alternative for the Pb potential. We also tried a Ca MEAM
potentential \cite{Kim2015}, but found that we obtain negative excess
volumes and excess entropies for some GBs, which seems unreasonable
for fcc materials, indicating that the potential is not suitable for
the simulation of GBs.

\subsection{Finding GB phases with the $\gamma$-surface method and
  calculation of GB excess properties}

\begin{figure}
  \includegraphics[]{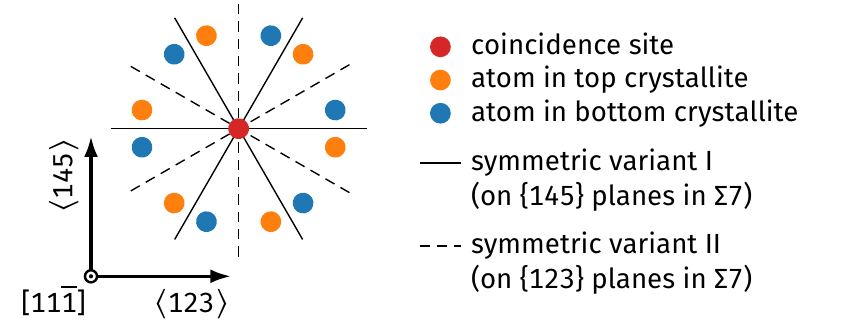}
  \caption{In $[11\overline{1}]$ tilt GBs in fcc, two symmetric
    variants exist. Here, part of the dichromatic pattern of a
    $\Sigma$7 GB is shown. The solid and dashed lines represent all
    possible symmetric GB planes (not considering translations). Due
    to the threefold symmetry of the $(11\overline{1})$ plane, all
    solid and dashed lines, respectively, are equivalent. In this
    work, we only investigate variant I.}
  \label{fig:symmetric-variants}
\end{figure}

\begin{figure}
  \includegraphics[]{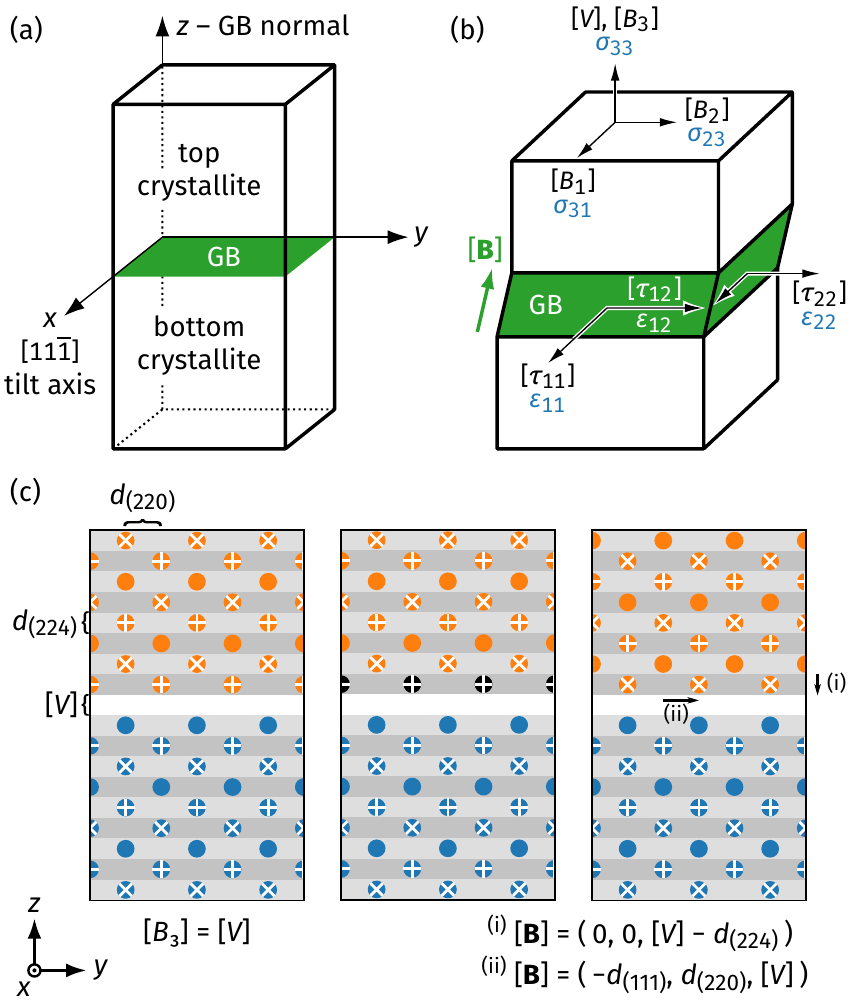}
  \caption{Geometry and excess properties of the bicrystals. (a) The
    convention used for the coordinate system. (b) The two
    crystallites are translated with regard to each other by a vector
    $[\mathbf{B}]$, which is specific to a GB phase. Due to the
    boundary conditions (lengths in $x$ and $y$ direction are fixed by
    the bulk phase and the system is free to expand in $z$ direction
    \cite{Frolov2012a}), excess stresses $[\tau_{11}]$, $[\tau_{22}]$,
    and $[\tau_{12}]$ occur, while the microscopic translations of the
    top crystallite lead to the excess volume $[V]$ and the excess
    shears $[B_1]$ and $[B_2]$. These excess properties couple to
    externally applied stresses $\sigma_{31}$, $\sigma_{23}$,
    $\sigma_{33}$ and strains $\varepsilon_{11}$, $\varepsilon_{22}$,
    $\varepsilon_{12}$ (blue text). (c) While the components $[B_1]$
    and $[B_2]$ are simple displacements, $[B_3]$ consists of both the
    excess volume $[V]$ and accounts for the shift by full
    crystallographic planes normal to the GB. Here, this is
    illustrated for a hypothetical $\Sigma3$ $[11\overline{1}]$
    GB. Different markers indicate the three different
    $(11\overline{1})$ planes and gray areas the interplanar distance
    $d_{(224)}$. The left side only has the $[V]$ component. (i)
    Removing a $\{112\}$ plane (black atoms in the middle), then leads
    to a shift downwards by $d_{(224)}$ (right side). (ii) This shift
    can alternatively be expressed by another shift that only changes
    the $[B_1]$ and $[B_2]$ components. The latter is always true in
    our GBs.}
  \label{fig:geometry}
\end{figure}

Next, we constructed the bicrystals. There are two sets of symmetric
grain boundaries for the [111] tilt axis
(Fig.~\ref{fig:symmetric-variants}). Here, we only consider variant I,
since two different GB phases have been identified before for
$\Sigma19$b $\langle111\rangle$ $\{178\}$ \cite{Meiners2020} and
$\Sigma37$c $\langle111\rangle$ $\{1\,10\,11\}$
\cite{Frommeyer2022}. For the symmetric variant II, only one GB phase
seems to exist \cite{Meiners2020a, LenaAg, SabaZoo}. We thus chose the
symmetric $\Sigma$ boundaries listed in Table~\ref{tab:bicrystals}. We
follow the convention from Ref.~\cite{Frolov2012a}, where $x$
corresponds to the tilt axis, $y$ to its orthogonal direction inside
the GB plane, and $z$ to the GB normal [Fig.~\ref{fig:geometry}(a)].

When searching for GB phases, bicrystals are joined together at the
desired GB plane and the microscopic degrees of freedom [translations
$[\mathbf{B}]$, Fig.~\ref{fig:geometry}(b)] are sampled
($\gamma$-surface method). We only sample $[B_1]$ and $[B_2]$ since
this can always be made equivalent to a full $[\mathbf{B}]$ vector by
addition of DSC vectors in our case [Fig.~\ref{fig:geometry}(c)].
Typically it is also necessary to consider inserting/removing partial
fcc planes at the GB in order to discover all relevant GB structures
\cite{Frolov2013, Hickman2017, Zhu2018a}. This can be expressed via
the parameter
\begin{equation}
  \label{eq:excess-n}
  [n] = \frac{N}{N_\text{plane}} \mod 1,
\end{equation}
where $N$ corresponds to the number of atoms in the bicrystal and
$N_\text{plane}$ to the number of atoms in a plane of the fcc
structure that is parallel to the GB. For the relevant $\Sigma19$b and
$\Sigma37$c GB phases, however, such search has found that all
defect-free GB phases have $[n] = 0$, i.e., no partial fcc planes
\cite{Meiners2020,Frommeyer2022}. We therefore assume that this is
true for the other $[11\overline{1}]$ tilt GBs and used the simple
$\gamma$-surface method. We verified the assumption of $[n] = 0$ by MD
simulations with open surfaces at high temperature \cite{Frolov2013}
for some example cases.

The GB excess properties were defined and calculated as described by
Frolov and Mishin \cite{Frolov2012, Frolov2012a}, except for the
microscopic, translational degrees of freedom $[\mathbf{B}]$, whose
calculation is described in Ref.~\cite{Frommeyer2022} and in
Supplemental Fig.~\ref*{fig:finding-B} \cite{suppl}. Detailed
definitions are also provided later in the paper together with the
results. Structures were visualized with \textsc{ovito}
\cite{Stukowski2010}. Raw data is available in the companion dataset
\cite{Brink2022zenodo}.

\begin{table}
  \caption{List of the bicrystalline samples used to construct
    symmetric $\Sigma$ tilt GBs and to search for GB phases.}
  \label{tab:bicrystals}
  \begin{ruledtabular}
  \begin{tabular}{lccll}
    CSL type & tilt axis & misorientation & \multicolumn{2}{c}{GB planes} \\
    \colrule
    \rule{0pt}{10pt}%
    $\Sigma$13b & $[11\overline{1}]$ & \ang{27.80}
                & $(\overline{7}2\overline{5})$ & $(\overline{7}5\overline{2})$  \\
    $\Sigma$7   & $[11\overline{1}]$ & \ang{38.21}
                & $(145)$ & $(415)$  \\
    $\Sigma$49  & $[11\overline{1}]$ & \ang{43.57}
                & $(2\,11\,13)$ & $(11\,2\,13)$  \\
    $\Sigma$19b & $[11\overline{1}]$ & \ang{46.83}
                & $(178)$ & $(718)$  \\
    $\Sigma$37c & $[11\overline{1}]$ & \ang{50.57}
                & $(1\,10\,11)$ & $(10\,1\,11)$  \\
  \end{tabular}
  \end{ruledtabular}
\end{table}

\subsection{Excess free energy}
\label{sec:methods:excess-free-energy}

In the present work, we are interested in the stability of GB phases
and GB phase transitions. In pure materials, such phase transitions
can occur under externally applied stress or strain, or with changing
temperature. The GB phase transitions can be predicted by computing
the GB excess free energies of the different GB phases, which we
define here as \cite{Frolov2012, Frolov2012a}
\begin{equation}
    \label{eq:gibbs}
    \gamma = [U] - T[S] - \sigma_{33}[V] - \sum_{i=1,2} [B_i] \sigma_{3i},
\end{equation}
where $[U]$ is the excess internal energy, $[S]$ the excess entropy,
$\sigma_{ij}$ the externally applied stress tensor, and $[V]$ the
excess volume. Here, $[B_1]$ and $[B_2]$ are excess shears, which are
equal to the microscopic, translational degrees of freedom when no
macroscopic stresses or strains are applied
[Fig.~\ref{fig:geometry}(b)]. The excess volume $[V]$ is not
necessarily equal to $[B_3]$ as defined here [see
Fig.~\ref{fig:geometry}(c)], but only $[V]$ enters the free energy. In
contrast to Refs.~\cite{Frolov2012, Frolov2012a}, we relax the
formality of the bracket notation for notational simplicity, intending
them only as indicators of GB excess values. We define all of them as
intrinsic values by normalizing $[U]$, $[S]$, and $[V]$ by the GB
area. More details and definitions of the excess properties are
provided later in Sec.~\ref{sec:excess-props}. The free energy at
$T = \SI{0}{K}$ under applied stresses and strains can be obtained
directly in molecular statics by applying the given stress to the
system and computing $[U]$, $[V]$, and $[B_i]$.

The influence of temperature, however, cannot be calculated directly,
because the entropy is not accessible via simple molecular statics or
MD simulations. Since we are dealing with pure systems, the entropy is
a vibrational entropy and can be computed either via thermodynamic
integration \cite{Frenkel1984, Freitas2016, Freitas2018} or with the
quasi-harmonic approximation (QHA) \cite{Foiles1994,
  Freitas2018}. Here, we chose the latter method for reasons of
computational efficiency. Force constant matrices were computed with
the \texttt{dynamical\textunderscore{}matrix} command in
\textsc{lammps}, from which we then obtained the phononic
eigenfrequencies in real space. The free energy $F$ was approximated
by negelecting quantum-mechanical effects as

\begin{equation}
  \label{eq:qha-free-energy}
  F = k_BT \sum_{i=1}^{3N-3}\ln\frac{h\nu_i}{k_BT},
\end{equation}
where $\nu_i$ are the phononic eigenfrequencies excluding the three
zero-valued eigenvalues. This describes the systems modeled by MD,
which are Newtonian systems, but we found that including
quantum-mechanical effects (mostly zero-point vibrations and
Debye-like thermal expansion at low temperatures) barely influences
the GB free energies, especially at room temperature and above
\cite{Frommeyer2022}. The GB excess free energy was calculated by
subtracting the free energy of a defect-free fcc slab containing the
same surfaces and number of atoms as the sample with the
GB.\footnote{Note that the subsystem method described in
  Ref.~\cite{Freitas2018}, which would not require a reference system
  containing the exact same surfaces and number of atoms, only works
  for thermodynamic integration, but not for the QHA. This is because
  the calculation of force constant matrices for a subsystem
  introduces an artificial boundary with its own excess free energy.}
Raw data is available in the companion dataset \cite{Brink2022zenodo}.

\subsection{Corroborative MD simulations}

In addition to calculating $\gamma(\sigma_{ij}, T)$, we also verified
some example cases using MD simulations at elevated temperature, with
and without applied stress and strain. We used systems roughly of size
$6 \times 16 \times \SI{20}{nm^3}$ (10 unit cells in tilt axis
direction) and a time integration step of \SI{1}{fs}. Temperature was
controlled with a Nos\'e--Hoover thermostat. We typically ran the
simulations for up to \SI{40}{ns}, or until the expected GB phase
transition could be observed. Raw data is available in the companion
dataset \cite{Brink2022zenodo}.

For calculations probing the influence of the temperature or the
stress $\sigma_{33}$ normal to the GB, we used open boundaries in $y$
and $z$ direction, while using periodic boundaries in the tilt axis
direction ($x$). A barostat at \SI{0}{Pa} was applied in the periodic
direction.

For the influence of a tension
or compression in $y$ direction ($\varepsilon_{22}$), we kept the $y$
direction periodic and instead introduced open boundaries in the $x$
and $z$ directions. Strain was applied in the periodic direction after
scaling the system to the appropriate lattice constant for the target
temperature and simulation cell length was subsequently kept constant
in the periodic direction.

\begin{figure*}
    \centering
    \includegraphics[]{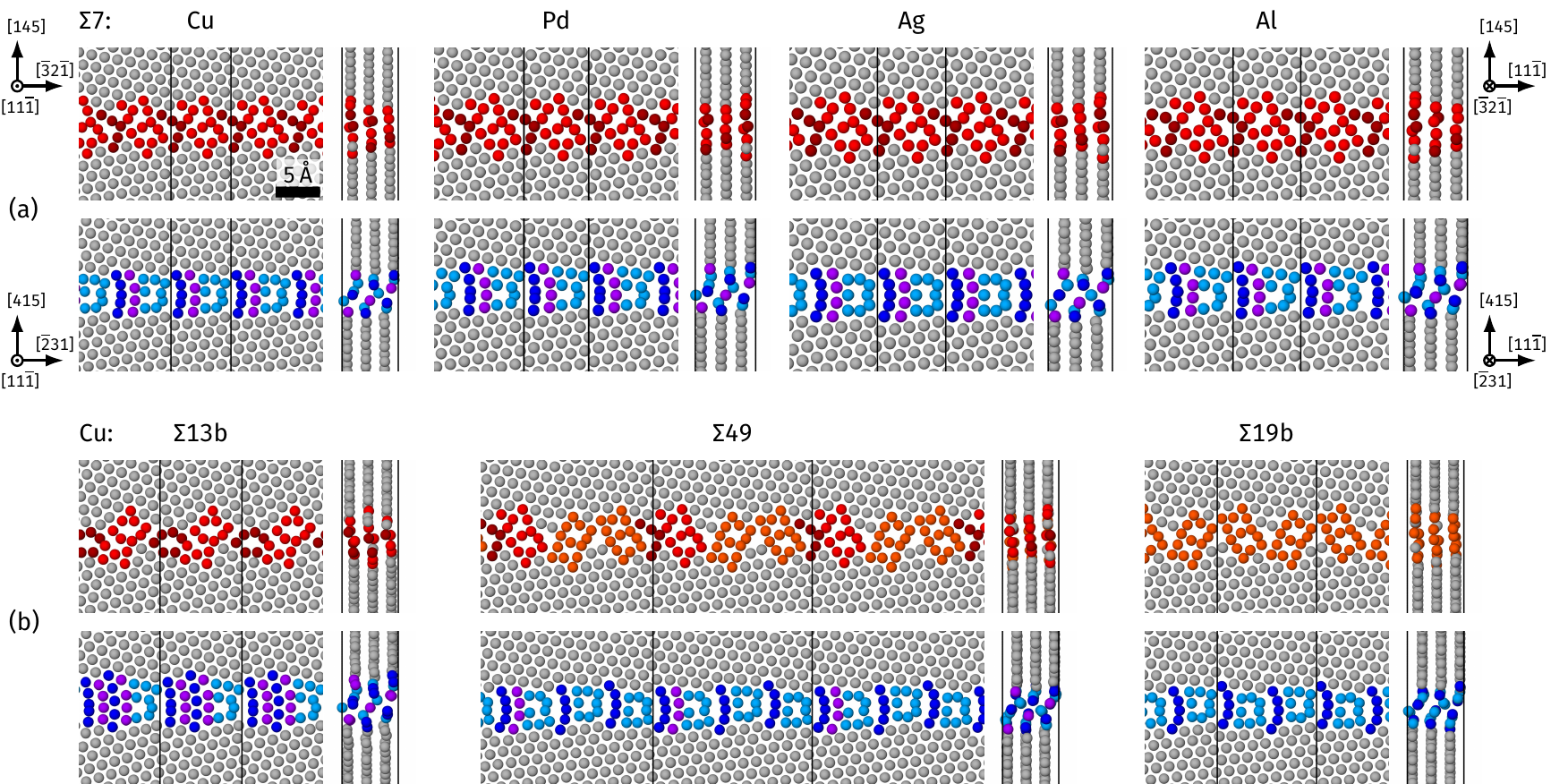}
    \caption{(a) GB phases of $\Sigma$7 tilt GBs in a selection of fcc
      metals. The top row shows the domino phase (red colors), the
      bottom row the pearl phase (blue colors). The scale bar is valid
      for all images and the axes indicate the crystal directions of
      the top and bottom crystallites, respectively. The axes on the
      left are for the top views, the axes on the right for the side
      views. The coloring serves only to highlight the structural
      motifs. (b) Different misorientations for copper GBs. The axes
      are equivalent to (a)---for the exact crystal directions see
      Table~\ref{tab:bicrystals}. Snapshots of the GB structures for
      all metals and misorientations can be found in Supplemental
      Figs.~\ref*{fig:snapshots-S13b}--\ref*{fig:snapshots-S37c}.}
    \label{fig:snapshots}
\end{figure*}

\subsection{Experimental sample preparation and STEM imaging}

In order to verify the simulations, we also experimentally
investigated the atomic structure of a near-$\Sigma49$ GB in copper in
addition to the already published experimental structures
\cite{Meiners2020, Frommeyer2022, SabaZoo}.

For this, a Cu thin film was deposited from a high purity (99.999\%)
Cu target on (0001)-oriented sapphire substrate by magnetron
sputtering. The deposition was performed at room temperature with a
radio frequency power supply at \SI{250}{W}, a background pressure of
\SI{0.66}{Pa}, and \SI{20}{sccm} Ar flow. We obtained a nominal film
thickness of \SI{600}{nm} with a deposition time of \SI{45}{min}. The
film was then annealed at \SI{400}{\celsius} for \SI{2}{h} within the
sputtering vacuum chamber.

We identified pure tilt high-angle grain boundaries using electron
backscattered diffraction imaging in a Zeiss Auriga scanning electron
microscope. In the next step, we lifted out a $\Sigma$49
$\langle111\rangle$ GB using a Thermo Fisher Scientific Scios2HiVac
dual-beam secondary electron microscope equipped with a Ga$^+$ focused
ion beam (FIB). A plane-view sample was extracted and attached to a Cu
grid. For the lamella thinning, an initial current of \SI{0.1}{nA} and
voltage of \SI{30}{kV} was used, reduced sequentially to \SI{7.7}{pA}
and \SI{5}{kV}. The FIB sample was then transferred to a
probe-corrected Thermo Fisher Scientific FEI Titan Themis 80-300
(scanning) transmission electron microscope. A high-brightness field
emission gun at an accelerating voltage of \SI{300}{kV},
semi-convergence angle of \SI{17}{mrad}, and probe current of
\SI{85}{pA} was used for imaging. The image was recorded with a
high-angle annular dark field (HAADF) detector (Fishione Instruments,
Model 3000) with a collection angle of 78 to \SI{200}{mrad}. An image
of 50 frames with $1024 \times \SI{1024}{px^2}$ with a dwell time of
\SI{2}{\micro s} and a step size of \SI{12.45}{pm} was registered and
overlaid using the drift compensated frame integration (DCFI)
method. The final image was optimized using second order polynomial
background correction, Butterworth, and Gaussian filters. The
misorientation between both grains was measured from the angles
between $\{220\}$ lattice planes of both grains, using an average of
at least ten different measurements.

\section{Structures in different fcc metals}

\begin{figure}
    \centering
    \includegraphics[]{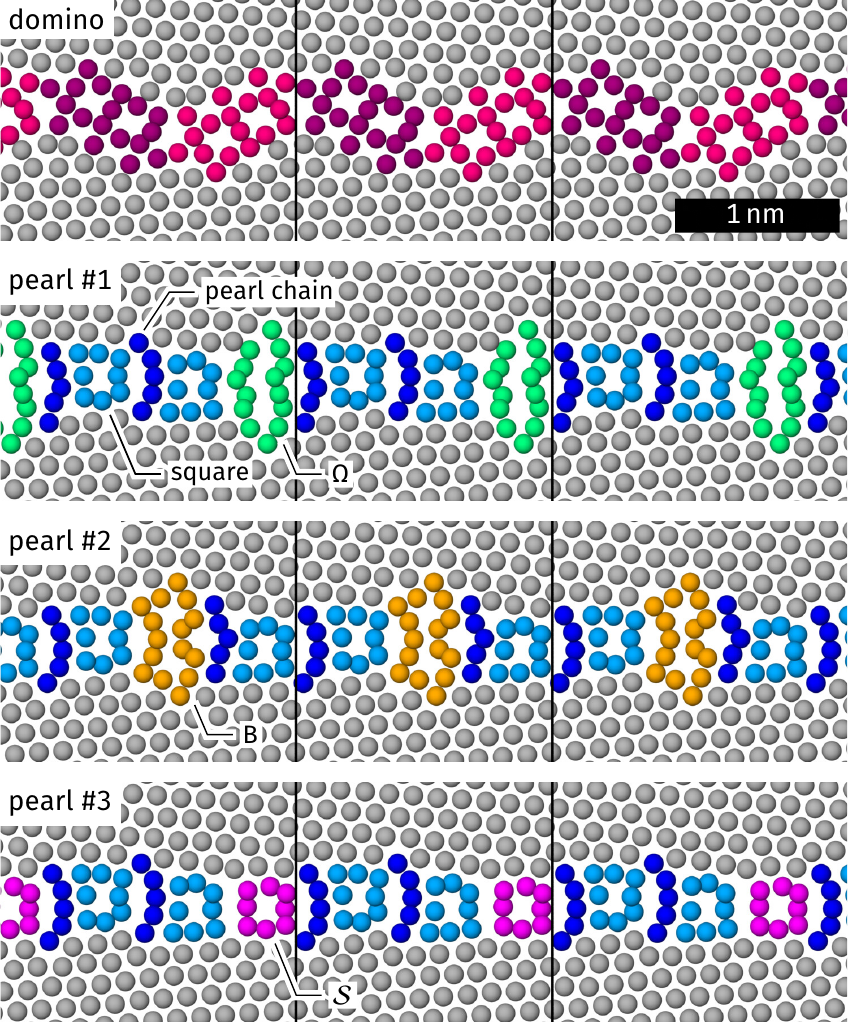}
    \caption{Snapshots of the domino phase and the three variants of
      the pearl phase in $\Sigma$37c. Here, the material is
      copper. The three pearl variants differ in their motifs. In
      addition to the typical squares and pearl chains, which are the
      same in all variants, we observe variations that we denominate
      with a letter: either an $\Omega$ motif (pearl \#1, green), a B
      motif (pearl \#2, orange), or an $\mathcal{S}$ motif (pearl \#3,
      pink) occurs. At $T = \SI{0}{K}$, the lowest-energy pearl
      variants are pearl~\#1 for Cu, Ag, and Au; as well as pearl~\#3
      for Ni, Pd, Al, and Pb. The energies of pearl~\#1 and pearl~\#2
      are typically very similar, except for Pb, where the pearl~\#2
      structure is mechanically unstable.}
    \label{fig:snapshots-S37c-pearls}
\end{figure}

We performed a computational structure search with Ni, Cu, Pd, Ag, Au,
Al, and Pb (M)EAM potentials for the GBs in
Table~\ref{tab:bicrystals}. We found that the possible structural
motifs are similar not only across different metals
[Fig.~\ref{fig:snapshots}(a) and Supplemental
Fig.~\ref*{fig:snapshots-S7}], but as well across different
misorientations from \ang{27.80} to \ang{50.57}
[Fig.~\ref{fig:snapshots}(b) and Supplemental
Figs.~\ref*{fig:snapshots-S13b}--\ref*{fig:snapshots-S37c}]. Indeed,
almost all of these motifs resemble the ``pearl'' and ``domino''
structures found previously in copper \cite{Meiners2020,
  Frommeyer2022}. In some cases, additional structures were found,
which will be discussed later.

Viewed from the $[11\overline{1}]$ tilt axis direction, the domino
phases consist of pairs of squares (light red motifs in
Fig.~\ref{fig:snapshots}), which are distorted and arranged
differently depending on the misorientation. From the side, the
$(11\overline{1})$ planes are approximately aligned, with offsets much
smaller than the interplanar spacing.

The pearl phases consist of a single square (light blue motifs in
Fig.~\ref{fig:snapshots}) separated by varying amounts of pearl chains
(dark blue or purple) when viewed from the $[11\overline{1}]$ tilt
axis direction. The $(11\overline{1})$ planes are shifted by
approximately half the interplanar spacing, in contrast to the domino
phases. Whereas the atomic structure of the domino phase seems to be
independent of the material [Fig.~\ref{fig:snapshots}(a), upper row],
slight changes can be observed in the pearl phase
[Fig.~\ref{fig:snapshots}(a), lower row]. In the $\Sigma$7 GBs, two
different pearl variants exist, which can be seen by comparing, e.g.,
Cu and Ag. The variant which occurs in Ag
[Fig.~\ref{fig:snapshots}(a)] and Pb (Supplemental
Fig.~\ref*{fig:snapshots-S7}) appears mirror symmetric in the
projection (called aligned pearl from here on), while the variant in
the other metals appears asymmetric (called sheared pearl from here
on). This minor difference can best be seen by inspecting the light
blue square motifs. Furthermore, additional motifs occur in the pearl
phase of the $\Sigma$37c GB (Fig.~\ref{fig:snapshots-S37c-pearls} and
Ref.~\cite{Frommeyer2022}), leading to three distinct pearl variants
(pearl \#1, \#2, and \#3).  Which of these variants has the lowest
energy depends on the material, as discussed later.

\begin{figure}
    \centering
    \includegraphics[]{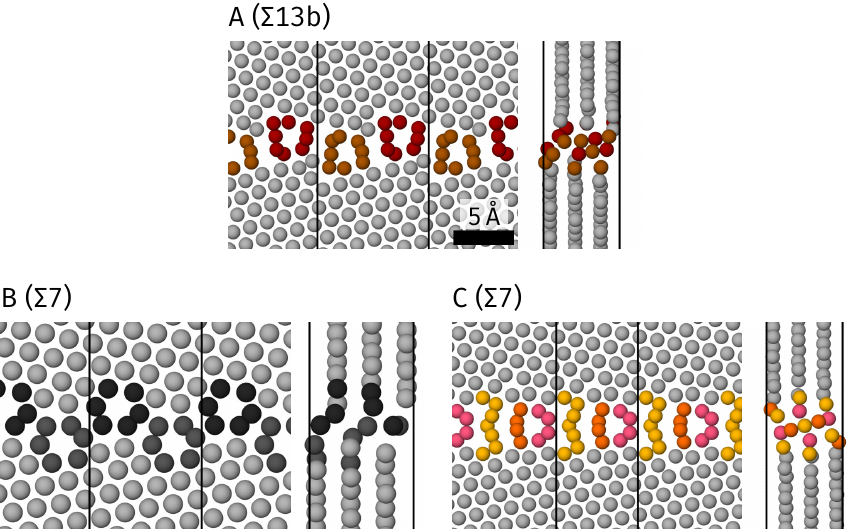}
    \caption{Snapshots of the A, B, and C phases. The material is
      copper for the A and C phases. The B phase only exists in
      lead. Snapshots for all materials are provided in
      Supplemental~Figs.~\ref*{fig:snapshots-S13b} and
      \ref*{fig:snapshots-S7}.}
    \label{fig:snapshots-ABC}
\end{figure}

For the $\Sigma13$b and $\Sigma7$ boundaries, some additional
low-energy structures occur, which we simply name A, B, and C (shown
in Fig.~\ref{fig:snapshots-ABC}). These look different from the pearl
and domino phases on visual inspection and are therefore listed
separately. We only consider structures that are thermodynamically
stable under some condition for at least one element and will later
provide a more quantitative analysis of the GB phases, but will
otherwise only discuss them where relevant. The A phase of the
$\Sigma$13b GB occurs in all metals except Au and Pb, but only has a
low energy compared to other GB phases in Al. In the $\Sigma$7 tilt
GBs, Pb has the B phase and the C phase is a higher-energy phase in
all metals.

\begin{figure}
    \centering
    \includegraphics{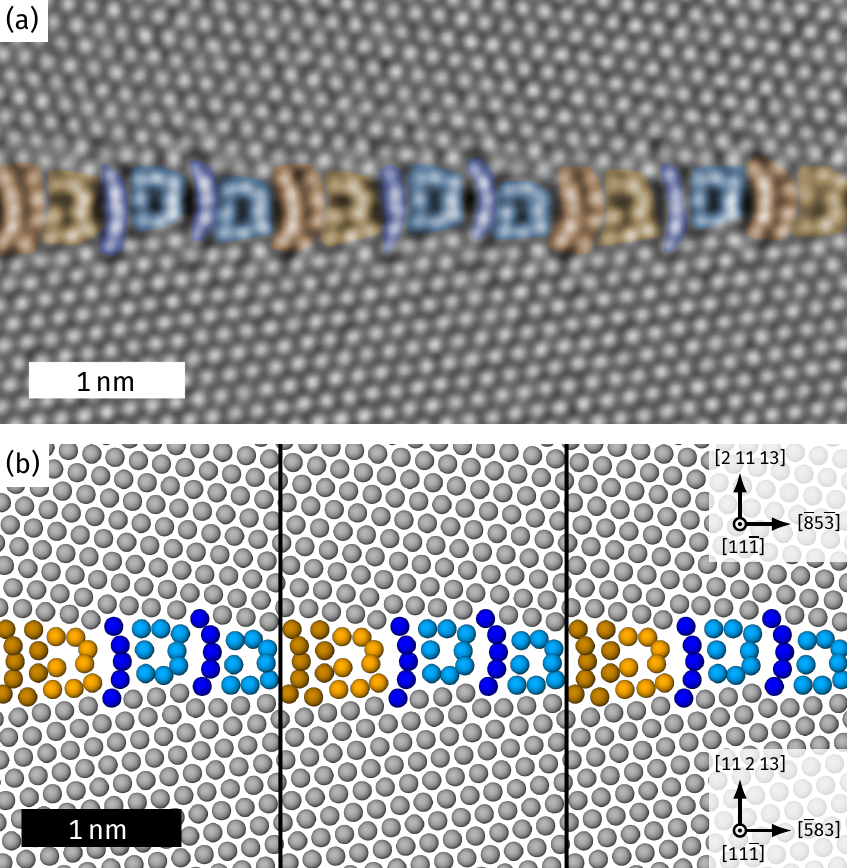}
    \caption{Comparison between a STEM image of a pearl structure in
      the $\Sigma$49 tilt GB in copper with simulations. (a)
      HAADF-STEM image of a $\Sigma49$ tilt GB (misorientation
      $\ang{43.3}\pm\ang{0.3}$). Note that there is a defect
      (disconnection) on the right side of the image, where only one
      blue square is located in between the yellow motifs and where
      the rightmost yellow motif is mirrored. An image of a longer
      stretch of the GB can be found in Supplemental
      Fig.~\ref*{fig:STEM-S49-full}. (b) It can be seen that the
      structure from the MD simulation is the same. The colors in
      these images highlight that the $\Sigma49$ GB consists of
      alternating motifs from the $\Sigma7$ GB (yellow) and the
      $\Sigma19$b GB (blue).}
    \label{fig:STEM}
\end{figure}

Due to the varying quality of empirical potentials---especially in the
case of GB structures which have not been included in the fitting
database for the respective potential---an independent validation
based on experiment or \textit{ab initio} methods is
desirable. Unfortunately, the unit cells of the high $\Sigma$ GBs are
too big to allow the required high-accuracy density-functional theory
(DFT) simulations, especially when one needs to avoid GB/surface
elastic interactions by including a sufficient amount of bulk
material. In tests we found that, e.g., the excess volume is very
sensitive to such effects. We thus limit ourselves here to a
comparison to STEM images obtained for Cu and Al. For Cu $\Sigma$19b
and $\Sigma$37c, see Refs.~\cite{Meiners2020, Frommeyer2022}. For
these GBs in Al, only the domino phase has been found to date
\cite{SabaZoo}. Additionally, Fig.~\ref{fig:STEM} shows an
experimental STEM image of a $\Sigma49$ pearl phase in copper. All of
the structures in the experiments listed above agree well with the
simulated structures. Because of the similarity of motifs across
misorientations and materials, we are confident that these structure
predictions are therefore quite reliable.

We note that the structures of the $[11\overline{1}]$ tilt GBs are
relatively complex compared to, for example, the more typical kite
structures in other (tilt) GBs \cite{Mills1992, Luo2011,
  Frolov2013}. Nevertheless, visual inspection already indicates
that---except for the special case of the B phase in
Pb---material-specific GB structures do not exist and that the
presented GB phases are universal in fcc metals. This raises the
question of how big the role of the physics and chemistry of a
specific material is and how much of that material-specific
information needs to be included in a model to reproduce them. We will
investigate this in the next section and then proceed to a more
quantitative comparison of the GB phases in terms of excess
properties.

\subsection{Geometry or material physics?}

\begin{figure}
    \centering
    \includegraphics[]{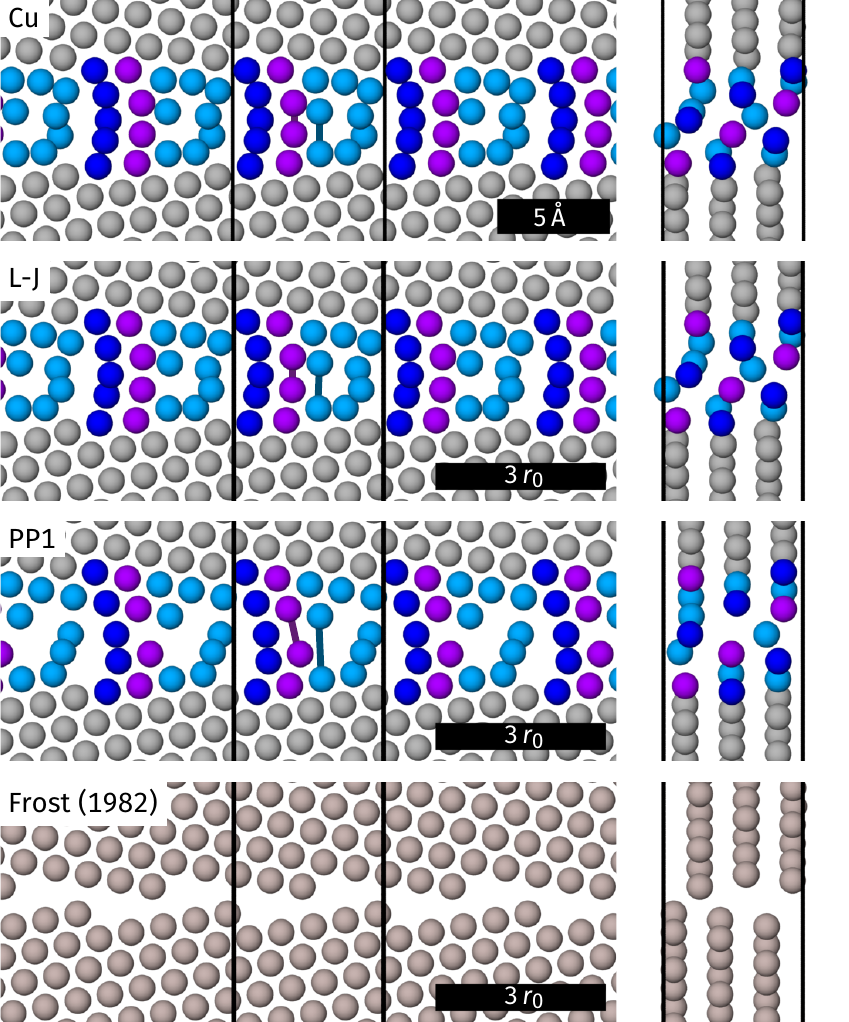}
    \caption{Visual comparison of the $\Sigma$7 pearl structure
      modeled using the copper EAM potential (Cu), a Lennard-Jones
      pair potential (L-J), and a next-neighbor pair potential
      (PP1). For comparison, the data from Frost et
      al.~(1982)~\cite{Frost1982} (rigid displacement of hard spheres)
      was also reproduced here, but does not exhibit any of the pearl
      or domino structures. The scale bar for the model potentials is
      given in reduced units of the equilibrium fcc bond length
      $r_0$. The axes are the same as in Fig.~\ref{fig:snapshots}(a).}
    \label{fig:snapshots-model-pots}
\end{figure}

As a first test, we used a Lennard-Jones pair potential as a
simplified, generalized model of a densely-packed metal. Physically,
pair potentials cannot reproduce the concept of bond order, i.e., the
strength of any interatomic bond simply depends on the bond length and
not on the atomic environment (such as for example coordination
number). By repeating the structure search with this pair potential,
we can nevertheless find the same structural motifs in the GBs (see
Fig.~\ref{fig:snapshots-model-pots} for the $\Sigma$7 pearl phase).

\begin{figure*}
    \centering
    \includegraphics{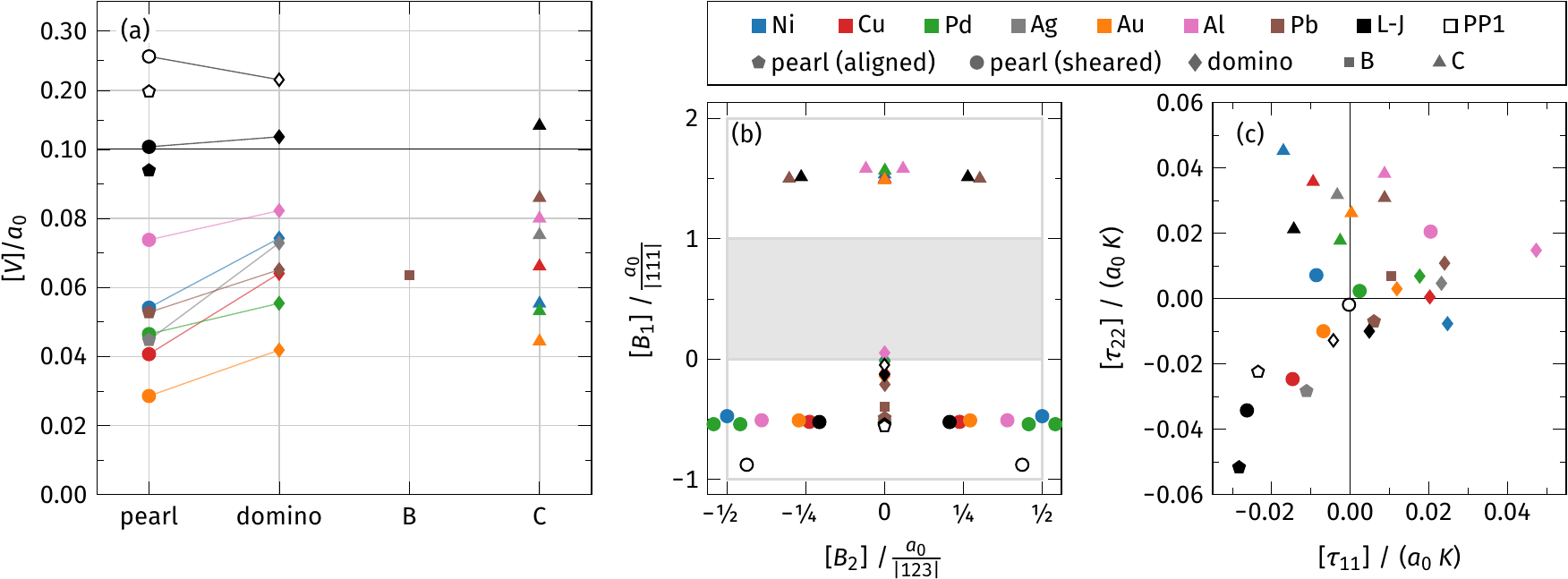}
    \caption{Excess properties of the $\Sigma$7 GBs. Colors of data
      points correspond to the material, shapes to the GB phase. (a)
      The normalized excess volumes of all relevant GB phases are
      shown. The connecting lines highlight the difference in excess
      volume between the lowest-energy pearl phase and the domino
      phase and reveal that the domino phases typically have higher
      excess volumes. Note that the upper part of the graph uses a
      different scale in order to be able to better discern the data
      from (M)EAM models. (b) The microscopic, translational degrees
      of freedom $[\mathbf{B}]$ of the $\Sigma$7 phases. The gray box
      shows the DSC unit cell, whose repetition is indicated by gray
      lines. The data are presented such that $[B_3] = [V]$ in order
      to show the data unambiguously. Data points lie outside of the
      projected DSC unit cell, because this unit cell is triclinic: To
      move the data points inside the unit cell, DSC vectors with an
      out-of-plane component are required, such that $[B_3] \neq [V]$
      (see Supplemental Fig.~\ref*{fig:excess-S7} for more details and
      different visualizations of this). (c) Two components of the
      excess GB stress normalized by lattice constant and bulk
      modulus. The general trend of tensile excess stresses for domino
      and compressive excess stresses for pearl can be seen. The C
      phase is distributed around $[\tau_{11}] = 0$ with tensile
      excess stresses in $[\tau_{22}]$. Data for all GBs are provided
      in Supplemental
      Figs.~\ref*{fig:excess-S13b}--\ref*{fig:excess-S37c}.}
    \label{fig:excess-vol-and-shears}
\end{figure*}

In the past, hypothetical GB structures have also been constructed
using the assumption of hard spheres due to the lack of realistic
interatomic potentials \cite{Frost1982}. We used our next-neighbor
pair potential to explore how realistic the results are under the
assumption of very short-ranged interactions. For this, we took the
set of all distinct GB structures obtained using all the other
potentials and reminimized them with the next-neighbor potential.
While both energy minimization using the next-neighbor pair potential
and rigid displacement of hard spheres (as performed by Frost et
al.~\cite{Frost1982}) lead to quite open GB structures, the former are
at least able to reproduce some of the motifs produced by
longer-ranged potentials (Fig.~\ref{fig:snapshots-model-pots}). The
two bonds drawn in Fig.~\ref{fig:snapshots-model-pots} are longer with
the next-neighbor pair potential than with the Cu and Lennard-Jones
potentials. This indicates that longer-ranged interactions and the
resulting local bond relaxations are crucial to describe GB structures
well. The more realistic, medium-ranged potentials predict an offset
between $(11\overline{1})$ planes. A look at the side view of the
structure modeled with the next-neighbor pair potential reveals that
there is almost no such offset. The hard sphere model interestingly
predicts the offset but not the other structural motifs.  This is not
necessarily the case for all misorientations, but the example of the
$\Sigma$7 tilt GB highlights that neither assumptions of next-neighbor
interactions nor of hard spheres will be sufficient to capture complex
GB structures and their excess properties.

We can conclude here that the GB phases are the result of the fcc
geometry and medium-ranged interatomic interactions, but that the GB
motifs are still densely packed, otherwise the Lennard-Jones potential
would not be able to model them. This purely visual inspection is
limited, however, which is why we will continue with an examination of
the excess properties for a more quantitative analysis.

\subsection{Excess properties}
\label{sec:excess-props}

A good definition of separate GB phases is that the corresponding
atomic structures have distinct excess properties. If the excess
properties are very close in value, we would rather define the
structures as defective or as microstates of a GB phase
\cite{Zhu2018a, Meiners2020, Frommeyer2022, Ahmad2023}. Furthermore,
the excess properties influence the GB thermodynamics
(Eq.~\ref{eq:gibbs}) and are therefore important quantities. We will
thus now discuss the individual excess properties introduced in
Sec.~\ref{sec:methods:excess-free-energy} in detail for our GB phases.

Figure~\ref{fig:excess-vol-and-shears}(a) shows the normalized excess
volumes $[V]/a_0$ of all $\Sigma$7 GB structures [data for other GBs
are shown in Supplemental
Figs.~\ref*{fig:excess-S13b}(a)--\ref*{fig:excess-S37c}(a)] with
\begin{equation}
  \label{eq:excess-vol}
  [V] = \frac{V_\text{GB} - N \Omega}{A_\text{GB}},
\end{equation}
where $V_\text{GB}$ is the volume of a region of the simulation cell
containing a GB (but no surfaces), $N$ is the number of atoms in that
region, $\Omega$ is the atomic volume in a defect-free fcc phase, and
$A_\text{GB}$ is the area of the GB. The normalization by the fcc
lattice constant $a_0$ at $T = \SI{0}{K}$ makes these volumes unitless and
comparable between materials.

In general, domino phases have higher excess volume than pearl phases
as indicated by the lines in Fig.~\ref{fig:excess-vol-and-shears}(a).
This trend is reproduced by the Lennard-Jones potential, but not by
the next-neighbor pair potential. As already visible in the snapshots
and as generally expected, this indicates that the relaxation inside
the GB is influenced by several neighbor shells.

Figure~\ref{fig:excess-vol-and-shears}(b) shows the microscopic,
translational degrees of freedom $[\mathbf{B}]$ of the $\Sigma 7$ GB,
which are the relative rigid-body displacements between the two
crystallites (data for other GBs are shown in Supplemental
Figs.~\ref*{fig:excess-S13b}--\ref*{fig:excess-S37c} and an
illustration of the concept can be found in Fig.~\ref{fig:geometry}).
$[\mathbf{B}] = 0$ represents the case when coincidence sites in the
dichromatic pattern actually overlap (which is how the dichromatic
pattern is typically plotted), while $[\mathbf{B}] \neq 0$ means that
no coincidence sites exist in the dichromatic pattern, i.e.,
$-[\mathbf{B}]$ represents the shift required to obtain coincidence
sites (see also Supplemental Fig.~\ref*{fig:finding-B}). Due to this,
$[\mathbf{B}]$ vectors are equivalent if they can be obtained by
adding or subtracting DSC vectors. The components $[B_1]$ and $[B_2]$
are also called excess shears and enter the GB excess free energy by
coupling to externally applied shear stresses (Eq.~\ref{eq:gibbs}).
Due to the symmetry of the present bicrystals with symmetric tilt GBs,
$[B_2]$ and $-[B_2]$ are degenerate states of the same GB phases. This
is not the case for $[B_1]$, where $-[B_1]$ corresponds to switching
the top and bottom crystallite.

A typical feature of the pearl phases is that the $(11\overline{1})$
planes of the abutting crystallites are not aligned, but shifted by
approximately a half-plane in tilt-axis direction. This is described
by $[B_1]$. All pearl variants are united by this half-plane
shift. The domino phases, in contrast, are characterized by
$[B_1] \approx 0$. The shift $[B_2]$ parallel to the GB plane has two
different behaviors in the case of the $\Sigma7$ GB. Both domino
(diamond symbols) and the aligned pearl variant (pentagon symbol) have
a fixed value, while the sheared pearl (circles) has
material-dependent values, i.e., exhibits some excess shear. The
difference in this shift describes the amount of shear that the pearl
motifs undergo. An aligned pearl variant was thus found in Ag and Pb
in the $\Sigma7$ GBs, and universally in the $\Sigma19$b GBs. The
$\Sigma13$b pearl phases are mostly aligned, except for a small
asymmetry in Ag, Pb, and the Lennard-Jones model, which we do not
indentify as a separate pearl variant. For $\Sigma49$ and $\Sigma37$c,
only sheared pearl variants exist.

Figure~\ref{fig:excess-vol-and-shears}(c) and
Supplemental~Figs.~\ref*{fig:excess-S13b}(f)--\ref*{fig:excess-S37c}(f)
show the normalized excess GB stresses $[\tau_{ij}] / (a_0K)$ with
\begin{equation}
  \label{eq:excess-stress}
  [\tau_{ij}] = \frac{\sigma_{ij}^\text{GB}V_\text{GB}}{A_\text{GB}},
\end{equation}
where $\sigma_{ij}^\text{GB}$ corresponds to the average of the
relevant stress tensor component in the region containing the GB. This
definition allows the easy calculation of strain energy via
$\int\! A_\text{GB} [\tau_{ij}] \mathrm{d}\varepsilon$ (cf.\
Eq.~\ref{eq:appl-strain-integral}), meaning that the excess stresses
are expressed in \si{J/m^2}. Here, the $[\tau_{11}]$ excess stress
acts along the tilt axis and $[\tau_{22}]$ along its orthogonal
direction within the GB plane [see also
Fig.~\ref{fig:geometry}(b)]. The normalization by the ground-state fcc
lattice constant $a_0$ and the bulk modulus $K$ makes these stresses
unitless and comparable between materials.

\begin{figure*}
    \centering
    \includegraphics[width=\linewidth]
                    {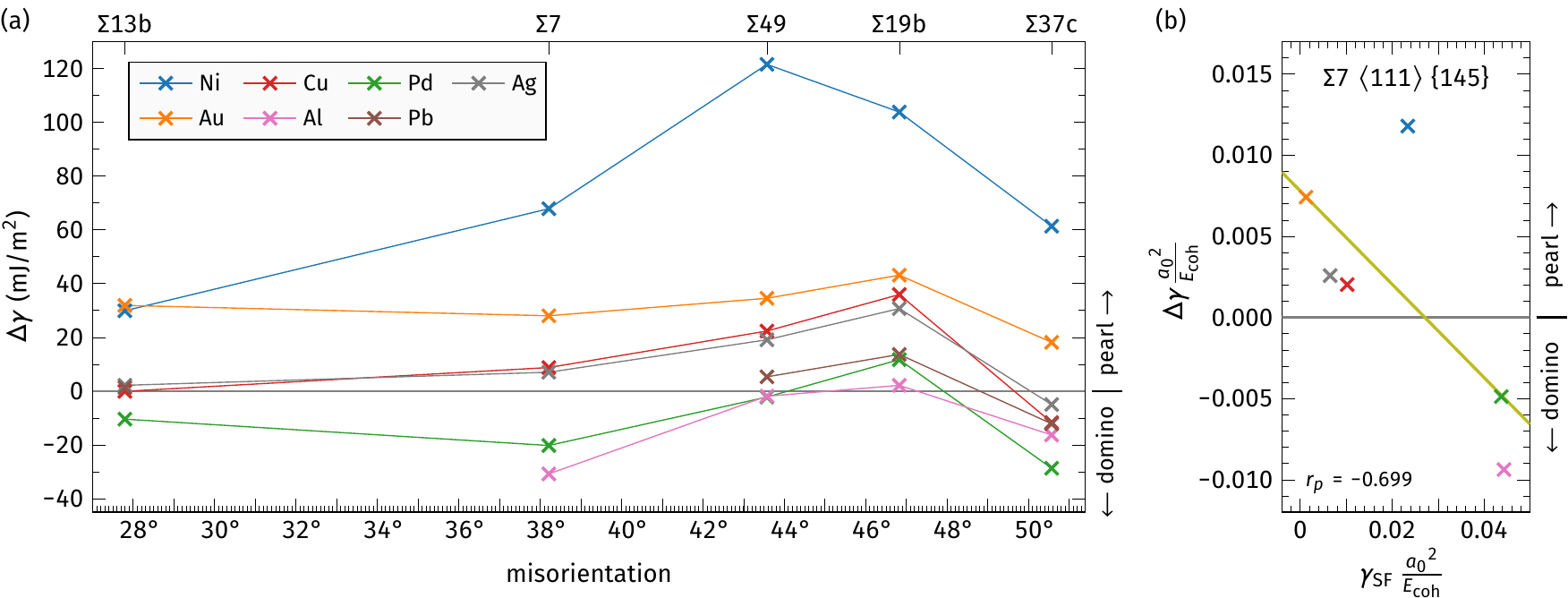}
    \caption{Ground state GB energy difference
      $\Delta\gamma = \gamma_0^\text{domino} - \gamma_0^\text{pearl}$.
      Positive values mean pearl is stable, negative values that
      domino is stable. These values are for $T = \SI{0}{K}$ and the
      difference between the lowest-energy pearl variant and the
      lowest-energy domino variant is used. For $\Sigma$13b, Al is
      omitted and for $\Sigma$7 Pb is omitted. This is due to other
      low-energy phases occuring that cannot be classified
      unambiguously as domino or pearl. (a) GB energy difference as a
      function of misorientation. The lines are guides for the
      eye. (b) Correlation between the normalized stacking-fault
      energy $\gamma_\text{SF}$ and $\Delta\gamma$. The normalization
      ensures that the plotted values are unitless and comparable
      between different materials. The line represents a linear
      regression of the data points and the Pearson correlation
      coefficient $r_p$ is reported. More data and the same plot
      without normalization is provided in Supplemental
      Fig~\ref*{fig:correlation-SFE-nonorm}.}
    \label{fig:ground-state-descriptors}
\end{figure*}

There is a tendency for $[\tau_{11}] < 0$ and $[\tau_{22}] < 0$ for
pearl and $[\tau_{11}] > 0$ and $[\tau_{22}] > 0$ for domino. Similar
trends are observed in all GBs. That means that the pearl phase with
lower excess volume is under compression in the in-plane directions of
the GB, while the domino phase with higher excess volume is under
tension. However, there are several exceptions, such as Al, where the
GBs generally tend more towards tensile stresses. The next-neighbor
pair potential is once again unable to capture this trend. We will
therefore not discuss it any further. The excess stresses of the pearl
variants for $\Sigma$37c GBs are very similar within the same
material, with the exception of Al [see Supplemental
Fig.~\ref*{fig:excess-S37c}(f)], indicating again that they are
closely related.

Supplemental Figs.~\ref*{fig:excess-S13b}--\ref*{fig:excess-S37c} show
that the A, B, and C phases are characterized by a $[B_1]$ offset of
half an interplanar spacing (similar to pearl), but positive
$[\tau_{22}]$ (similar to domino), which is why we label them as
individual GB phases. We make the distinction between A and B/C
because they appear at different CSL boundaries, and differentiate B
and C in Pb by their different values of excess shear $[B_2]$ and
excess stress $[\tau_{22}]$.

\section{Thermodynamic stability}

\subsection{Ground state stability}

For each of the investigated tilt GBs we showed that at least two GB
phases (domino and pearl) exist. In order to evaluate which of these
GB phases will actually occur, we have to investigate their
thermodynamics. We first consider the case of $T = \SI{0}{K}$ and no
externally applied stress or strain. Here, the ground state GB free
energy is
\begin{equation}
  \label{eq:gb-energy}
  \gamma_0 = [U] = \frac{E_\text{GB} - N E_0^\text{fcc}}{A_\text{GB}},
\end{equation}
where $E_\text{GB}$ is the potential energy of a region of the
simulation cell containing a GB (but no surfaces), $N$ is the number
of atoms in that region, $E_0^\text{fcc}$ is the ground-state energy
per atom of the defect-free fcc phase, and $A_\text{GB}$ is the area
of the GB. Note that the internal energy is equal to the potential
energy in this section, because we simulate a classical, Newtonian
system at zero temperature, i.e., with zero kinetic energy. All GB
energies are plotted in Supplemental Fig.~\ref*{fig:gb-energy}.

Figure~\ref{fig:ground-state-descriptors}(a) and Supplemental
Fig.~\ref*{fig:correlation-SFE-nonorm} show the energy difference
$\Delta\gamma = \gamma_0^\text{domino} - \gamma_0^\text{pearl}$ at
$T = \SI{0}{K}$ between the domino and pearl phases. In many cases,
the pearl phase is stable, with the highest energy difference being
observed for the $\Sigma19$b GB, except for Ni (highest energy
difference for $\Sigma49$). The $\Sigma19$b GB often also has the
lowest overall GB energy, although for Pd, Al, and Pb the $\Sigma7$ GB
has a lower energy (Supplemental Fig.~\ref*{fig:gb-energy}). The
domino phase is stable in the $\Sigma37$c GB in many materials, and
seems to also become more favorable again towards lower misorientation
angles. Apart from domino and pearl, the A, B, and C phases are
usually high-energy GB phases. Exceptions are the A phase in Al, which
is quite close in GB energy to the pearl phase, and the B phase in Pb,
which is the lowest-energy GB phase in $\Sigma7$. The C phase only
plays a role at higher temperatures, as discussed later in
Sec.~\ref{sec:temperature}.

The Lennard-Jones potential, in contradiction to the more realistic
(M)EAM potentials, always predicts that the domino phase is stable
[Supplemental Fig.~\ref*{fig:correlation-SFE-nonorm}(a)]. This
indicates that the Lennard-Jones potential is not a good
``generalized'' model for most metals: While it is able to capture
structural motifs and their excess properties qualitatively, the
material-specific physics play a bigger role in the relative
thermodynamic stability of the GB phases. This is also important for
the evaluation of early simulation work on GB structures
\cite{Sutton1983}: The present results suggest that it is likely that
such simulations predicted the wrong ground state because they used
pair potentials instead of more realistic material models.

For the (M)EAM potentials, we tried to find a simple predictor of the
relative GB phase stability as a function of the material properties,
and focus here on the stacking-fault energy $\gamma_\text{SF}$. This
is a tempting quantity, because both GBs and stacking faults are
planar defects and because $\gamma_\text{SF}$ is indirectly related to
the energy of coherent twin GBs. If we normalize both the
stacking-fault energy of the metal and the GB energy difference
$\Delta\gamma$ between domino and pearl by the respective fcc lattice
constant and cohesive energy, we find that low stacking-fault energies
tend to be associated with a preference for the pearl phase and high
stacking-fault energies with a preference for the domino phase
[Fig.~\ref{fig:ground-state-descriptors}(b) and Supplemental
Fig.~\ref*{fig:correlation-SFE-nonorm}]. This trend is not very
convincing, however, since for example nickel deviates quite strongly
from the correlation. The data for the $\Sigma13$b and $\Sigma37$c GBs
is also quite scattered, exhibiting a tendency for an increased
stability of domino. A further relation between stacking-fault energy
and structure is that the high stacking-fault-energy metals Ni, Pd,
and Al have lower energies for the pearl \#3 variant in $\Sigma37$c,
while the low stacking-fault-energy metals Cu, Ag and Au (as well as
the Lennard-Jones potential) favor pearl \#1 or \#2. Yet, this
relation is undermined by Pb, whose stacking-fault energy is low, but
whose ground-state pearl variant is \#3. Other material properties
have even weaker or no correlation with $\Delta\gamma$.

Ultimately, the relative energies of GB phases have to be computed
with sophisticated models (such as EAM potentials or DFT) and simple
rules based on bulk properties will be incomplete. This is especially
true since the GB energy differences are often small and on the order
of \SI{10}{mJ/m^2}, although they can be as high as approximately
\SI{120}{mJ/m^2} [Fig.~\ref{fig:ground-state-descriptors}(a)].

\subsection{Stress-dependence of the free energy}

\begin{figure}[t!]
    \centering
    \includegraphics{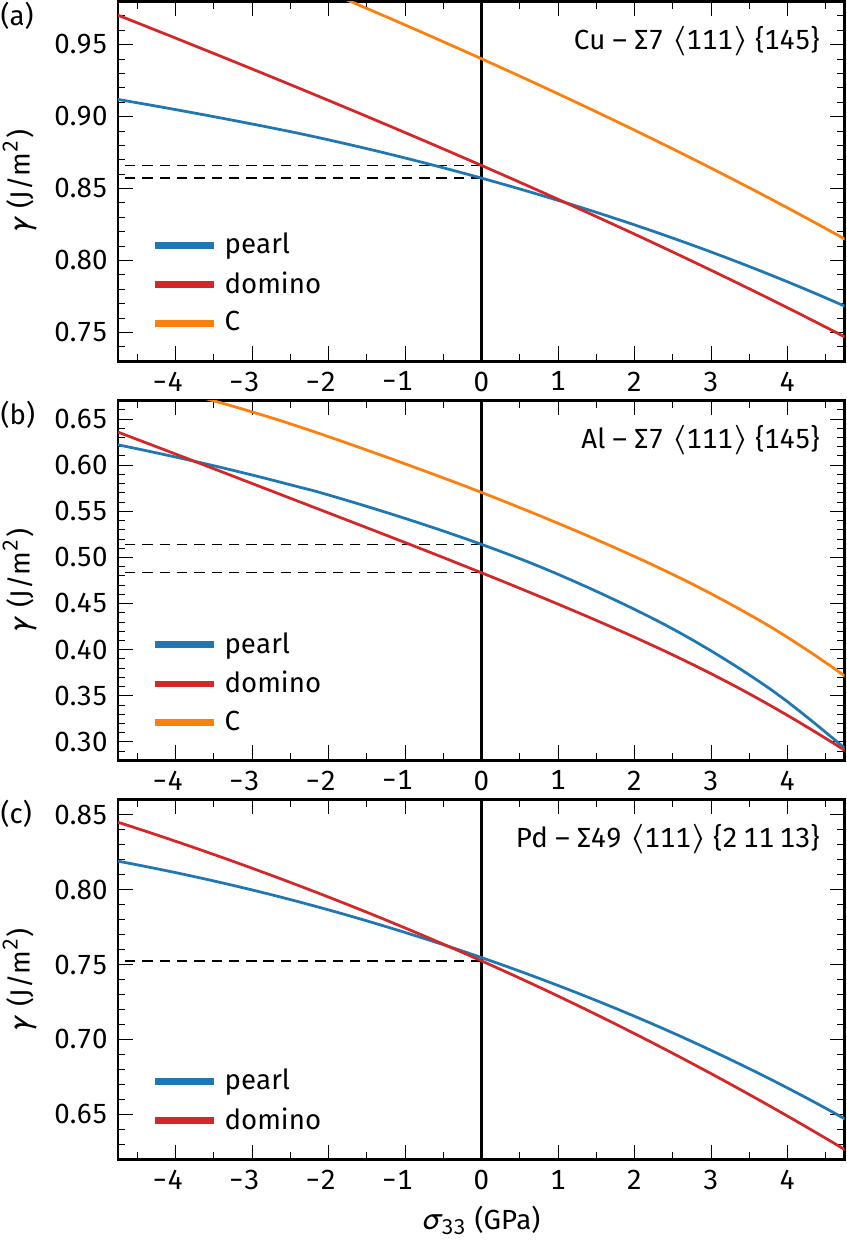}
    \caption{GB free energies as a function of an applied stress
      $\sigma_{33}$ normal to the GB for some selected GBs. (a) In
      copper $\Sigma7$ GBs, the domino phase becomes stable under
      tension. (b) The Al potential also predicts that pearl becomes
      stable under very high compressive stresses. Furthermore, it
      exhibits some changes of the curvature of $\gamma$ for the pearl
      phase, which are associated with small rearrangements of the
      atomic motifs, leading to the prediction that pearl can
      additionally become stable at very high tensile stresses.  This
      is likely due to the Al potential being described by cubic
      splines (the change of curvature representing a transition to
      another polynomial in the spline). Other potentials do not show
      such behavior, suggesting that this is most likely an unwanted
      and unphysical feature of the potential at high
      stresses/strains. (c) The Pd $\Sigma49$ GB represents an
      interesting case due to the low stresses required to transition
      between pearl and domino. Data for all GBs is available in
      Supplemental
      Figs.~\ref*{fig:sigma33-S13b}--\ref*{fig:sigma33-S37c}.}
    \label{fig:free-energy-stress}
\end{figure}
\begin{figure}[t!]
    \centering
    \includegraphics{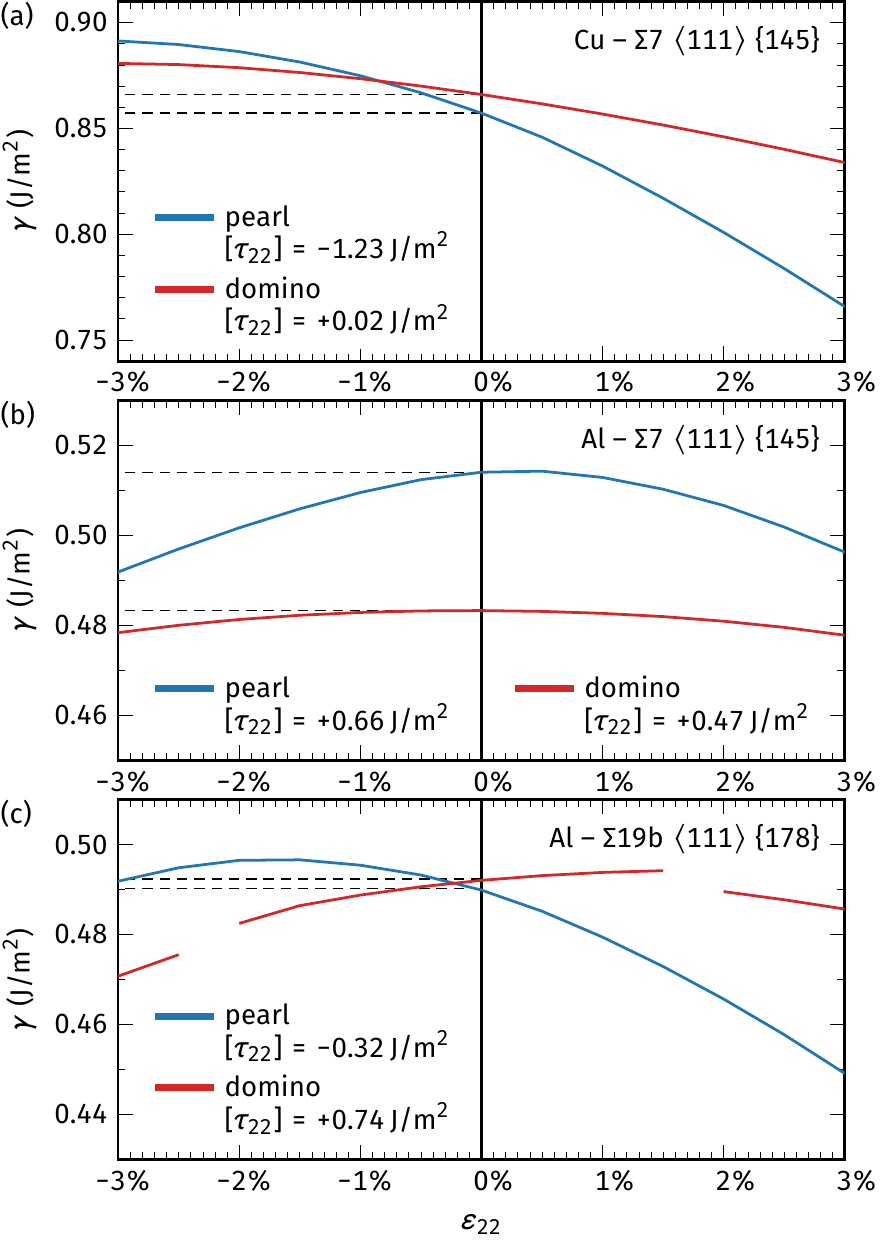}
    \caption{GB free energies as a function of an applied strain
      $\varepsilon_{22}$. (a) The domino phase in the $\Sigma$7 GB in
      Cu can be stabilized under compression. (b) In Al, domino remains
      stable independent of the applied strain. This is due to both GB
      phases having positive excess stresses of similar magnitude,
      meaning that their excess free energy change due to strain is
      comparable.  (c) In the $\Sigma19$b GB, however, the excess
      stresses of domino and pearl in Al have opposite signs and a
      phase transition is possible. The gaps in the curve represent
      points at which the atoms slightly rearrange within the domino
      motifs, leading to small jumps in the excess free energy. Data
      for all GBs is available in Supplemental
      Figs.~\ref*{fig:epsilon22-S13b}--\ref*{fig:epsilon22-S37c}.}
    \label{fig:free-energy-strain}
\end{figure}

The excess free energy at $T = \SI{0}{K}$ under applied stresses
$\sigma_{33}$, $\sigma_{23}$, and $\sigma_{31}$ can be described as
\cite{Frolov2012a}
\begin{equation}
  \label{eq:appl-stress}
  \gamma = [U] - \sigma_{33}[V] - \sum_{i=1,2}[B_i]\sigma_{3i}
\end{equation}
in the case of tilt GBs [see Fig.~\ref{fig:geometry}(b) for a sketch
of the stresses and strains]. The excess free energy under applied
strain $\varepsilon_{22}$ can be calculated by integrating the work
done by the excess stress \cite{Frolov2012a} as
\begin{equation}
  \label{eq:appl-strain-integral}
  \gamma = \gamma_0\frac{A_0}{A_\text{GB}} +
  \frac{1}{A_\text{GB}}
  \int_0^{\varepsilon_{22}}
  A_\text{GB}(\varepsilon) \, [\tau_{22}](\varepsilon) \, \mathrm{d}\varepsilon,
\end{equation}
where $A_0$ is the GB area before deformation. The calculation for
$\varepsilon_{11}$ and $\varepsilon_{12}$ is equivalent. Because the
additional work is converted fully into excess internal energy, we can
simply express the GB free energy as
\begin{equation}
  \label{eq:appl-strain}
  \gamma = [U],
\end{equation}
without need for the integration.

Because the domino phase universally has a higher excess volume than
the pearl phase [Fig.~\ref{fig:excess-vol-and-shears}(a)], it stands
to reason that it can be stabilized by a tensile stress $\sigma_{33}$
normal to the GB, while the pearl phase can be stabilized by a
compressive stress.  Figure~\ref{fig:free-energy-stress} and
Supplemental Figs.~\ref*{fig:sigma33-S13b}--\ref*{fig:sigma33-S37c}
show that this is the case, but that the required stresses are usually
very high (on the order of gigapascals) and sometimes exceed the range
of values that were investigated. One interesting exception is for
example the Pd $\Sigma$49 GB [Fig.~\ref{fig:free-energy-stress}(c)],
where the ground state energy difference between the domino and pearl
phases is close to zero. We thus validated this case by MD simulations
at $T = \SI{900}{K}$ and $\sigma_{33} = \pm\SI{1}{GPa}$, starting once
from a pearl phase and once from a domino phase. In less than
\SI{8}{ns}, the systems under compression transitioned to the pearl
phase and the systems under tension to the domino phase, as expected
(see Supplemental Fig.~\ref*{fig:md-sigma33-S49-Pd}).

Another interesting excess property is the excess stress $[\tau_{22}]$
in the GB plane, since it indicates if the GB phase is under
compression or tension. Since typically $[\tau_{22}] < 0$ for pearl
and $[\tau_{22}] > 0$ for domino, it seems reasonable that a
compressive $\varepsilon_{22}$ would favor domino, while tension would
favor pearl (see Eq.~\ref{eq:appl-strain-integral}).
Figure~\ref{fig:free-energy-strain} and Supplemental
Figs.~\ref*{fig:epsilon22-S13b}--\ref*{fig:epsilon22-S37c} show that
this is indeed a general trend. Exceptions are some GBs in Ni, Al, and
Pb. For those GBs, however, the value of $[\tau_{22}]$ is positive for
the pearl phase, suppressing a clear trend for a GB phase transition
under applied strain. For $\Sigma13$b GBs in Cu, Ag, or Pb; $\Sigma7$
Gbs in Cu or Ag; $\Sigma49$ GBs in Pd or Al; $\Sigma19$b GBs in Al;
and $\Sigma37$c GBs in Cu or Ag, the required strains for the GB phase
transitions are relatively low and could reasonably be observed
experimentally (Supplemental
Figs.~\ref*{fig:epsilon22-S13b}--\ref*{fig:epsilon22-S37c}). We tested
this by performing MD simulations for the $\Sigma7$ GB in Cu with
$\varepsilon_{22} = -2\%, +1\%$. We ran the simulations at
$T = \SI{900}{K}$ in order to accelerate the transition kinetics. A GB
phase transition can be observed after less than \SI{4}{ns}, obtaining
the domino phase under compression and the pearl phase under tension
(Supplemental Fig.~\ref*{fig:md-epsilon22-S7-Cu}).

In addition to externally applied stresses, the rigid-body
displacements between the two crystallites could also be determined by
restrictions on GB sliding in polycrystals or by the bonding to a
substrate in the thin film case \cite{Ahmad2023}. Then, however,
$[\mathbf{B}]$ in Eq.~\ref{eq:appl-stress} no longer corresponds to
the displacement, but the \textit{excess} displacement over the
defect-free crystal subject to the same stress $\sigma_{3i}$
\cite{Frolov2012, Frolov2012a}. Thus, knowledge of the resulting,
system-size-dependent stress state would be required to be able to
calculate the free energy, necessitating mesoscale modeling.

Finally, we found that the A, B, and C phases are not stable, except
for the A phase in Al, which becomes stable under compression normal
to the GB plane or under tensile $\varepsilon_{22}$, and the B phase
in Pb, which is the lowest-energy GB phase in $\Sigma7$ even without
applied stress or strain.

In summary, our simulations show that the GB phase transitions both
under stress and under strain can be well predicted by the excess
properties in the ground state. The GBs thus obey Le Chatelier's
principle and counteract the applied stresses and strains via GB phase
transitions.

\subsection{Temperature-dependence of the free energy}
\label{sec:temperature}

The excess free energy without applied stresses or strains is
\begin{equation}
  \label{eq:temperature}
  \gamma = [U] -T[S]
\end{equation}
for finite temperatures. Note that for $T > \SI{0}{K}$, the potential
energies $E$ in Eq.~\ref{eq:gb-energy} have to be replaced by
$\langle E \rangle$, which are the average total energies at the given
temperature. The excess entropy $[S]$ is not easily accessible to MD
simulations. The change of excess free energy
$\gamma(T) = \gamma_0 + \Delta\gamma(T)$ can instead be calculated
using either thermodynamic integration \cite{Frenkel1984, Freitas2016,
  Freitas2018} or from the phonon eigenfrequencies using the QHA
\cite{Foiles1994, Freitas2018}. We chose the latter due to the lower
computational demands. In previous works, both methods were found to
give equal results \cite{Freitas2018, Frommeyer2022}.

\begin{figure}[t!]
    \centering
    \includegraphics[width=\linewidth]{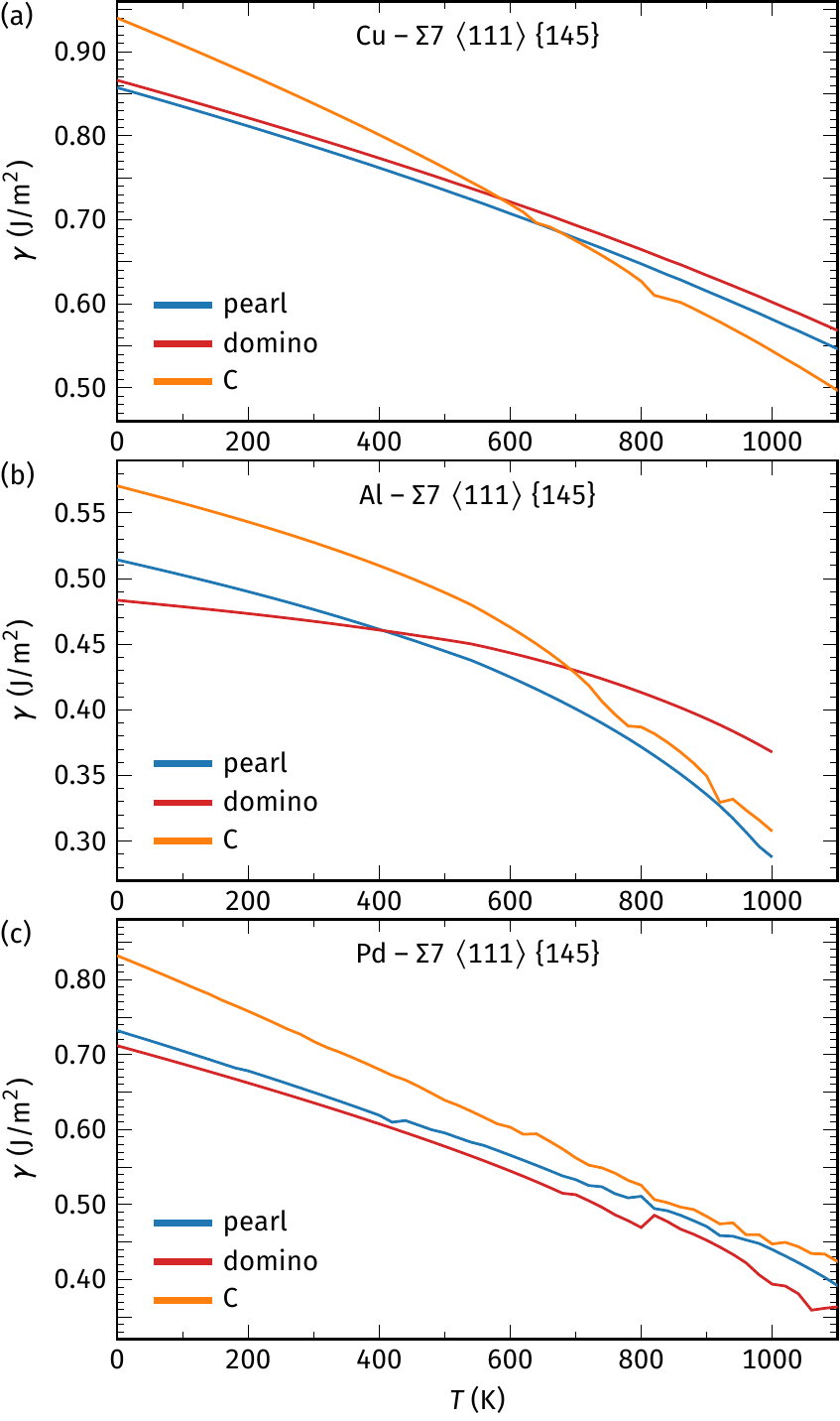}
    \caption{GB free energies as a function of temperature for some
      selected $\Sigma7$ GBs. (a) In copper, the pearl phase is
      stable, until the high-energy, high-temperature C phase becomes
      stable at around \SI{650}{K}. (b) In Al, domino transitions to
      pearl at around \SI{400}{K}, (c) while in Pd the domino phase
      remains stable at all temperatures. In the latter cases, the C
      phase is always metastable. The noise in the data is a result of
      numerical issues with the interatomic potentials, see
      Appendix~\ref{sec:numerical-issues}. Data for all GBs is
      available in Supplemental
      Figs.~\ref*{fig:temperature-S13b}--\ref*{fig:temperature-S37c}.}
    \label{fig:free-energy-temperature}
\end{figure}

In most cases, the pearl phase remains stable over the whole
temperature range (Fig.~\ref{fig:free-energy-temperature} and
Supplemental
Figs.~\ref*{fig:temperature-S13b}--\ref*{fig:temperature-S37c}). If
the domino phase is stable at low temperatures ($\Sigma$13b: Pd, Al;
$\Sigma7$: Pd, Al; $\Sigma$49: Al; $\Sigma37$c: Cu, Pd, Ag, Al, Pb) it
usually transforms to pearl at higher temperatures, except for
$\Sigma$7 Pd (domino is always stable). In $\Sigma13$b Pb, as well as
$\Sigma19$b Ag and Pb, pearl transforms to domino at higher
temperatures. In $\Sigma49$ Pd, both phases have approximately equal
free energy.
Some limited experimental data is available for copper
(Refs.~\cite{Meiners2020, Frommeyer2022} and Fig.~\ref{fig:STEM}) and
supports the modeling. In Al, only the domino phase has been found to
date \cite{SabaZoo}. That contradicts the data obtained with the Al
potential for the $\Sigma19$b GB (but not the others). The predicted
energy differences between domino and pearl are small, however, which
complicates the comparison to experiment: Either the relative
stability of the GB phases is predicted incorrectly by the potential
or small residual strains in the experiments, which were performed on
thin films, could potentially destabilize the pearl phase.

The only non-domino/non-pearl phases that become stable with
increasing temperature are the A phase in Al $\Sigma13$b GBs and the C
phase in Cu and Ag $\Sigma7$ GBs
(Fig.~\ref{fig:free-energy-temperature} and Supplemental
Figs.~\ref*{fig:temperature-S13b} and \ref*{fig:temperature-S7}). In
general, the investigated GBs mostly exhibit domino and pearl phases,
with A, B, and C being the exceptions.

In order to furthermore exclude the existence of additional GB phases
and to validate the QHA calculations, we ran MD simulations for up to
\SI{40}{ns} with the GBs in contact with open boundaries at elevated
temperatures \cite{Frolov2013} for some example cases. We chose high
temperatures to enable GB phase transitions on MD timescales. No
additional GB phases were discovered. For the $\Sigma7$ GBs, we found
that Ni ($T = \SI{1300}{K}$) and Al ($T = \SI{900}{K}$) transition to
the pearl phase at high temperatures independent on the starting
structure, as expected (Supplemental
Figs.~\ref*{fig:md-temperature-S7-Ni} and
\ref*{fig:md-temperature-S7-Al}). The Al sample contained many defects
after the heat treatment. Both Cu ($T = \SI{1100}{K}$) and Ag
($T = \SI{1100}{K}$) quickly transitioned to the pearl phase if
starting from domino and slowly nucleated the C phase, also as
expected (Supplemental Figs.~\ref*{fig:md-temperature-S7-Cu} and
\ref*{fig:md-temperature-S7-Ag}). In the latter case, however, C and
pearl phases often appear to coexist, hinting that the C phase is some
variant of the pearl phase. The only case that could not confirm the
QHA calculations was the Pd $\Sigma7$ GB: neither systems containing
pearl, nor containing domino would undergo phase transitions when
heated to \SI{900}{K} (Supplemental
Fig.~\ref*{fig:md-temperature-S7-Pd}). This is likely due to the small
free energy differences between the GB phases and possibly a result of
the low mobility of the phase junction between the two GB phases
\cite{Meiners2020}. Apart from the $\Sigma7$ GBs, we also annealed a
Ni $\Sigma37$c GB at \SI{1000}{K} (Supplemental
Fig.~\ref*{fig:md-temperature-S37c-Ni}). The QHA calculations predict
that the pearl \#3 variant is stable at low temperatures and pearl \#1
and \#2 at high temperatures, the latter GB phases having almost equal
free energies [Supplemental~Fig.~\ref*{fig:temperature-S37c}(a)]. We
could indeed observe this transition between pearl variants, although
the result at high temperature still contained several pearl \#3
motifs. This is not surprising, since these structures should most
likely be treated as microstates of a pearl GB phase
\cite{Frommeyer2022}.

Due to the rough tendency of pearl being stable at higher
temperatures, it is tempting to treat the pearl structure as a
high-entropy GB phase. But while it is universal that domino has
higher excess volumes and thus couples to $\sigma_{33}$, a higher
entropy of the pearl phase cannot be observed in the majority of
cases. When pearl is stable over the whole temperature range, the
slope of $\gamma(T)$ is often similar for pearl and domino, indicating
approximately equal excess entropies. In pure materials, the excess
entropy is vibrational and results from the GB phonon modes. This
means that the GB vibrations are therefore quite material dependent.

\subsection{Latent heat and order of the GB phase transition}

\begin{figure}
    \centering
    \includegraphics[width=\linewidth]{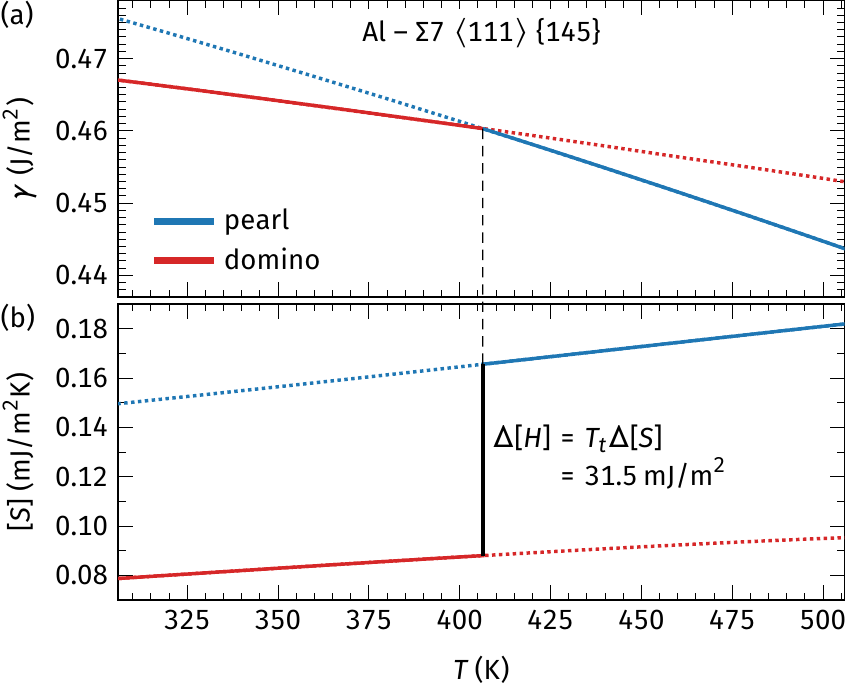}
    \caption{Order of the GB phase transition for the Al $\Sigma7$
      GB. The solid lines indicate the stable GB phase. (a) In
      equilibrium, the domino phase is stable at low temperatures,
      while the pearl phase is stable above $T \approx
      \SI{406}{K}$. (b) This results in a discontinuity of the
      entropy, resulting in a latent heat of
      $\Delta[H] = \SI{31.5}{mJ/m^2}$. This value approximately
      corresponds to the difference in ground state GB energies
      $\gamma_0^\text{pearl} - \gamma_0^\text{domino} =
      \SI{30.8}{mJ/m^2}$.}
    \label{fig:latent-heat}
\end{figure}

Finally, we will shortly discuss the order of the GB phase
transition. According to the Ehrenfest classification, first order
phase transitions are those that have a discontinuity in the first
derivative of the relevant thermodynamic potential. Alternatively,
first order phase transitions have a latent heat. In bulk materials,
these definitions are virtually equivalent and the latent heat is the
difference in enthalpy $\Delta H = \Delta G + T \Delta S$ of the
phases, with $G$ being the Gibbs free energy and $S$ the entropy. By
definition, it is $\Delta G = 0$ at the transition point $T_t$,
allowing us to write $\Delta H = T_t \Delta S$. The entropy of the
system in equilibrium is discontinuous at $T_t$ for $\Delta S \neq 0$
due to the change from one phase to another with different
entropy. This means that the phase transition is also of the first
order according to Ehrenfest because the entropy of a phase is the
first derivative of the free energy ($S = - \partial G / \partial T$)
at constant pressure. Equation~\ref{eq:temperature} with
$\Delta \gamma = 0$ leads to a similar result for GBs, namely
$\Delta [H] = \Delta [U] = T_t \Delta [S]$ for $\sigma_{3i} = 0$ (no
work is done by the system on its surroundings without externally
applied stresses, only heat is exchanged). The excess entropy can be
calculated as
\begin{equation}
  \label{eq:entropy1}
  [S] = -\frac{\mathrm{d}\gamma}{\mathrm{d}T}
        + \sum_{i,j=1,2} \bigl([\tau_{ij}] - \delta_{ij}\gamma\bigr)
          \frac{\mathrm{d}\varepsilon_{ij}}{\mathrm{d}T},
\end{equation}
(see Eq.~14 of Ref.~\cite{Frolov2012a}, noting that we already
normalize $[S]$ by the GB area) where the strains $\varepsilon_{ij}$
correspond to the thermal expansion of the grains, which result in
work being done by the grain boundary against the expansion, even
without externally applied stress. (Note that this work term does not
come into play when defining the latent heat as
$\Delta [H] = \Delta [U]$, because the phase transformation takes
place at constant $T = T_t$ and thus constant $\varepsilon_{ij}$. It
is only required to calculate $[S]$.)  For cubic systems, we can
replace $\mathrm{d}\varepsilon_{ij}/\mathrm{d}T$ with the isotropic
thermal expansion coefficient $\alpha_T$ and simplify to
\begin{equation}
  \label{eq:entropy2}
  [S] = -\frac{\mathrm{d}\gamma}{\mathrm{d}T}
        + \alpha_T \bigl( [\tau_{11}] + [\tau_{22}] - 2\gamma \bigr).
\end{equation}
We ignore the weak temperature dependence of the excess stresses in
further calculations. Figure~\ref{fig:latent-heat} shows an example
for the $\Sigma7$ GB in Al. At the transition temperature of around
\SI{405}{K}, the excess free energy of the system in equilibrium
changes slope and the entropy is discontinuous, resulting in a finite
latent heat. Equivalent results would be obtained for the other GB
phase transitions. The GB phase transitions in this work are thus
first order phase transitions.

\section{Summary and conclusion}

Simulations using (M)EAM potentials reveal that a range of high-angle,
symmetric $[11\overline{1}]$ tilt GBs in fcc metals exhibit mainly two
GB phases, here called domino and pearl. We found that the domino and
pearl phases, respectively, have comparable structures and
thermodynamic excess properties across different $\Sigma$ boundaries
with misorientations from \ang{27.8} to \ang{50.6} and for all of the
seven investigated fcc metals.  Indeed, the structures seem universal
enough to be modeled using simple pair potentials, although
medium-range interatomic interactions are required to recover the
trends of the excess properties. The thermodynamic stability as a
function of stress, strain, or temperature, however, is more specific
to the material:
\begin{itemize}
\item In many cases the pearl phases are stable at $T = \SI{0}{K}$ and
  if they are not, they often become stable at higher temperatures due
  to their higher excess entropy. This is not universal however,
  suggesting that the GB vibrations are material dependent.
\item There does not seem to be a clear predictor of which
  GB phase is the ground state, although there is some weak
  correlation with the stacking-fault energy, with low stacking-fault
  energy favoring pearl.
\item The domino phases have higher excess volumes in almost all
  cases, meaning that they are stabilized by tension applied normal to
  the GB, while the pearl phases are stabilized under compression.
\item The pearl phases tend to exhibit negative excess stresses in the
  GB plane ($[\tau_{11}]$ and $[\tau_{22}]$), while the domino phases
  exhibit positive excess stresses. This predicts their thermodynamic
  stability quite well: compression in the GB plane stabilizes domino,
  while tension stabilizes pearl, opposite to the previous case of
  stress applied normal to the GB.
\item The required stresses for GB phase transformation can exceeed
  \SI{5}{GPa} and would be unlikely to occur in real materials. In
  some cases, however, stresses below \SI{1}{GPa} or even close to
  zero, as well as strains below 1\%, are sufficient. We confirmed
  some of the latter cases with MD simulations.
\end{itemize}
While there are always some exceptions to the above rules, the present
results suggest that GB structures and phases discovered for one
material are likely generalizable to a whole class of materials (in
this case fcc metals). It remains to be seen if that is a feature of
densely-packed metals or if it is also true in, e.g., covalently or
ionically bonded materials. The addition of alloying elements is also
likely to lead to more material-specific GB thermodynamics.

\begin{acknowledgments}
  This project has received funding from the European Research Council
  (ERC) under the European Union's Horizon 2020 research and
  innovation programme (Grant agreement No.~787446; GB-CORRELATE).

  \textit{Author contributions:} T.B. and G.D. designed the
  study. T.B. performed and analyzed the simulations and wrote the
  initial manuscript draft. L.L. performed the STEM investigation and
  H.B. synthesized the corresponding thin film. G.D. supervised the
  project, contributed to discussions, and secured funding for T.B.,
  L.L., and H.B. via the ERC advanced grant GB-CORRELATE. All authors
  participated in the preparation of the final manuscript.\\
\end{acknowledgments}

\appendix

\section{Numerical issues of the interatomic potentials}
\label{sec:numerical-issues}

The calculation of free energies in the QHA depends on the force
constant matrix, which is closely related to the Hessian of the
potential energy. This requires that the interatomic potential is
smooth up to the second derivative, otherwise the resulting free
energy is noisy. In MD simulations, only first derivatives of the
potential energy are used and any noise is averaged out due to the
natural thermal fluctuations of the dynamic simulation, making
numerical issues with the potential unnoticeable. In the present work,
we found that the Pd, Au, and Al potentials result in noisy free
energy data in our QHA calculations.

For Al, this is not due to the formalism of the potential, but a
consequence of limited precision of the tabulated potential data in
the file distributed via the NIST Interatomic Potentials Repository
\cite{NIST-IPR}. In \textsc{lammps}, EAM potentials are
tabulated---typically with dense sampling of data points---in text
files and interpolated during simulation. In current numerical
simulations, floating point numbers are usually represented using the
IEEE 754 double-precision format, which corresponds to a precision of
at least 15 significant digits in decimal representation. The text
files used to produce the Al potential file contain fewer significant
digits. We tested the potential by using the $\Sigma19$b domino GB
phase and isotropically straining the simulation cell, while recording
the potential energy (Supplemental Fig.~\ref*{fig:Al-pot}). Taking
numerical derivatives of the potential energy, we obtain smooth curves
up to the first derivative, but see noise from the second derivative
on. This is not surprising, since numerical derivatives are
particularly sensitive, even to small noise. We then recovered the
original nodal points, reproduced the potential, and constructed a
tabulated EAM potential file with full machine precision (see
companion dataset \cite{Brink2022zenodo} for details and the potential
file). This leads to smooth results up to the third derivative of the
potential energy (Supplemental Fig.~\ref*{fig:Al-pot}) while
preserving the properties of the potential (we verified this for the
bulk properties listed in Table~\ref{tab:mater-prop-other}, as well as
the excess properties of the GB phases).

Additionally, the construction of the Al potential from cubic splines
leads to changes of curvature of the GB free energy [visible for
example in Fig.~\ref{fig:free-energy-stress}(b) at high stresses or in
Fig.~\ref{fig:free-energy-temperature}(b) at around \SI{550}{K}]. This
indicates a shortcoming of the fitting with cubic splines to a limited
reference database. It seems that the flexibility of representing the
EAM functions with splines would require more strained reference data
points in order to correctly model highly strained systems.

For Pd and Au, the first and second derivatives of the potential
energy are sufficiently smooth (Supplemental
Fig.~\ref*{fig:Pd-Au-pots}). The problem comes from the cutoff
function in these cases. It appears that the cutoff functions are
relatively abrupt. This works well for the defect-free crystalline
structures, where interatomic distances are usually smaller or larger
than the cutoff distance. In the case of our GBs, however, bond
lengths around the cutoff distance can appear and certain bond lengths
can cross this distance during thermal or mechanical straining. As
depicted in Supplemental Fig.~\ref*{fig:Pd-Au-pots}, this can lead to
discontinuities in the forces and thus jumps and/or noise in the
second derivatives. It is nontrivial to improve the existing
potentials without changing their properties and we therefore use the
unchanged potentials here.

\end{document}